\documentclass{aastex631}

\usepackage{CJK}

\usepackage{graphicx}	
\usepackage{amsmath}	
\usepackage{amssymb}	
\usepackage{float}
\usepackage{graphicx}
\usepackage{subfigure}
\usepackage{multirow}
\usepackage{enumerate}
\usepackage{color}
\usepackage{amsmath}
\usepackage{tabularx}
\usepackage{booktabs}
\usepackage{array}
\usepackage[flushleft]{threeparttable}

\submitjournal{ApJS}

\shorttitle{Blind search for 1179 star clusters}
\shortauthors{Chi et al.}

\graphicspath{{./}{figures/}}

\begin{document}
\begin{CJK*}{UTF8}{gbsn}

\begin{sloppypar}
\title{Blind Search of The Solar Neighborhood Galactic Disk within 5kpc :  1,179 new Star clusters found in Gaia DR3}

\correspondingauthor{Feng Wang, Hui Deng, Zhongmu Li}
\email{fengwang@gzhu.edu.cn, denghui@gzhu.edu.cn, zhongmuli@126.com}

\author[0000-0001-7343-7332]{Huanbin Chi  ({\CJKfamily{gbsn}迟焕斌})}
\affiliation{Center for Astrophysics and Great Bay Center of National Astronomical Data Center, \\ Guangzhou University, Guangzhou 510006 , China}
\affiliation{School of Management and Economics, Kunming 650500, China}
\affiliation{Peng Cheng Laboratory, Shenzhen, 518000, China}

\author[0000-0002-9847-7805]{Feng Wang ({\CJKfamily{gbsn}王锋})}
\affiliation{Center for Astrophysics and Great Bay Center of National Astronomical Data Center, \\ Guangzhou University, Guangzhou 510006 , China}
\affiliation{Peng Cheng Laboratory, Shenzhen, 518000, China}

\author[0000-0002-9847-7805]{Wenting Wang ({\CJKfamily{gbsn}王雯婷})}
\affiliation{School of Physics and Astronomy, Sun Yat-sen University Zhuhai Campus, Zhuhai 519082, China;}

\author[0000-0002-8765-3906]{Hui Deng ({\CJKfamily{gbsn}邓辉})}
\affiliation{Center for Astrophysics and Great Bay Center of National Astronomical Data Center, \\ Guangzhou University, Guangzhou 510006 , China}
\affiliation{Peng Cheng Laboratory, Shenzhen, 518000, China}

\author[0000-0002-0240-6130]{Zhongmu Li ({\CJKfamily{gbsn}李忠木})} 
\affiliation{Institute of Astronomy, Dali University,  Dali, 671003,  China}

\begin{abstract}
Studying open clusters (OCs) is essential for a comprehensive understanding of the structure and evolution of the Milky Way. Many previous studies have systematically searched for OCs near the solar system within 1.2 kpc or 20 degrees of galactic latitude. 
However, few studies searched for OCs at higher galactic latitudes and deeper distances.
In this study, based on a hybrid unsupervised clustering algorithm (Friends-of-Friends and pyUPMASK) and a binary classification algorithm (Random Forest), we extended the search region (i.e., galactic latitude $\left|b\right|\geq$ 20 $^\circ$) and performed a fine-grained blind search of Galactic clusters in Gaia DR3.
After cross-matching, the newly discovered cluster candidates are fitted using isochrone fitting to estimate the main physical parameters (age and metallicity) of these clusters. 
These cluster candidates were then checked using manual visual inspection. Their statistical properties were compared with previously exposed cluster catalogs as well. In the end, we found 1,179 new clusters with considerable confidence within 5kpc.

\end{abstract}


\section{Introduction} \label{sec:intro}
Galaxies or open clusters (OCs) are chemically homogeneous stellar populations that have the same age, the same kinematics (proper motion and radial velocity), and maintain approximately the same separation from us. 
OCs are ideal laboratories and powerful tools for studying star formation and evolution~\citep{Krumholz2019, Bossini2019}. For example, the vast majority of OCs are located near the Galactic plane and thus serve as excellent tracers of the recent formation history of the Galactic disk. 

Accurate determination of cluster membership is critical for the study of open clusters, as it directly impacts the estimation of their fundamental astrophysical parameters.
Most studies~\citep{CG2018,CG2018A,CG2019,CG2019A,CG2020,He2022ApJS1,CG2022} use an unsupervised machine learning algorithm, such as the Density-Based Spatial Clustering of Applications with Noise algorithm (DBSCAN).
DBSCAN could search for arbitrarily shaped clusters by adjusting two parameters, i.e., the neighborhood radius (Eps) and the density threshold (MinPts). However, a set of Eps and MinPts can only locate clusters with a specific distribution density. Therefore, searching for clusters with different member star densities with only a set of global Eps and MinPts results in a significant identification bias, leading to clusters being over-segmented or merging multiple clusters into one~\citep{Hunt2021,2020PhD}.

In addition to single clustering algorithms, hybrid algorithms have been effectively studied. \citet{Chi2022} (hereafter Paper I) proposed a hybrid method of pyUPMASK and Random Forest (RF) to identify potential OCs, and 46 reliable clusters were successfully re-authenticated out of 807 clusters removed after \citet{Li2022} identification, which proves that the hybrid presented method is effective. 
The hybrid method consists of 3 steps: the friends-of-friends (FoF) algorithm for rough clustering, members census, and OC identification by RF model.
The advantage of the FoF algorithm to group stars is that clustering considers a five-dimensional weighted parameter space of parallax, position, and velocity.
We investigated related algorithms that have been widely used in searching for star clusters and presented in Table~\ref{tab:list_OCs}. 


\begin{table}[htbp]
\centering
	\caption{List of main previously published SC catalogues.}
	
	\begin{threeparttable}
\begin{tabular*}{\hsize}{@{}@{\extracolsep{\fill}}p{5cm}cp{1cm}cp{1cm}lp{1cm}rp{2cm}@{}} 
	\toprule
		Works & OCs number & Catalog Name & Method (Based Data)&Type\\
		\midrule
	\cite{Dias2002}, \cite{Dias2012}, \cite{Kharchenko2013} & 3006   & MWSC  & Collected and Compiled(WEBDA et al.)& pre-Gaia clusters \\
	\cite{Liu_Pang2019} & 76 & LP  & FoF (Gaia DR2)  &  Gaia cluster \\
\cite{Ferreira2019}, \cite{Ferreira2020}, \cite{Ferreira2021}  & 62   & Ferreira Series  &Joint Analysis (Gaia DR2,EDR3)&Gaia clusters \\
\cite{He2021},  \cite{He2022apjs}, \cite{He2022ApJSb},
\cite{He2022ApJS1}  & 3157   & CWNU    & DBSCAN(Gaia DR2,EDR3)&Gaia clusters \\
\cite{Hao2020},
 \cite{Hao2021},
 \cite{Hao2022A},
 \cite{Hao2022}  & 4552 \tnote{1}  & Hao Series  & DBSCAN(Gaia DR2, EDR3)&Gaia clusters \\
  \cite{CG2018}, \cite{CG2018A}, \cite{CG2019}, \cite{CG2019A}, \cite{CG2020}, \cite{CG2020A}, \cite{CG2020B}, \cite{CG2022}  & 5305 \tnote{2}  & UBC   & DBSCAN and UPMASK(Gaia DR2,EDR3)&Gaia clusters \\
  \cite{Li2022}, \citet{Li2022b}  & 96   & LISC   & FoF (Gaia DR2, EDR3)&Gaia clusters \\
  \citet{Chi2022,Chi2022a}  & 129   &   & Hybrid Method (Gaia DR2, EDR3)&Gaia clusters \\
   \cite{Hunt2021}  & 41   & HR21  & HDBSCAN (Gaia DR2 ,EDR3)&Gaia clusters \\
	\citet{Vasiliev2021}  & 170   & GC  & GMMs (Gaia eDR3)&Globular clusters \\
	\citet{Dias2021}  & 1743   &   & Collected and Compiled (Gaia DR2)&Gaia clusters \\
	\citet{Tarricq2022}& 389   &   & Collected and HDBSCAN (Gaia EDR3)&Gaia clusters \\
	\citet{Jaehnig2021}& 431\tnote{3}   & XDOCC  & Collected and XDGMMs (Gaia DR2)&Gaia clusters \\
	\citet{Casado2021}& 30   & Casado  & Manual Mining (Gaia DR2)&Gaia clusters \\
	\citet{Bica2019}& 10978   & B19  & Collected (multi-band catalog)&Star Clusters, Associations and Candidates \\
	\citet{Torrealba2019}& 90   &   & Statistics  Modelling (Gaia DR1, DES, Pan-STARRS)&Star Clusters  \\
	\citet{Qin2021RAA}& 4   & QC  & Field Star Decontamination (Gaia DR2) & Gaia Clusters  \\
	\citet{sim2019}   & 207 & UPK  & Visual Search (Gaia DR2)        & Gaia Clusters  \\
		\bottomrule
	\end{tabular*}
	\end{threeparttable}
	\begin{tablenotes}
	\footnotesize
	\item{1} included 1930 previously
    known open clusters and compiled a
    catalogue of 3794 OCs based on Gaia EDR3. \item{2} series included 2017+1229 previously reported clusters.
    \item{3} included 420 previously found OCs.
    \\ GMMs means mixture Gaussian modelling clustering algorithm.  HDBSCAN is the Hierarchical Density-Based Spatial Clustering of Applications with Noise clustering algorithm.
    XDGMMs is a “top-down” technique, Extreme Deconvolution Gaussian Mixture Models.
	\end{tablenotes}
	\label{tab:list_OCs}
\end{table}

However, the search for open clusters is a long and challenging task ~\citep{Deb2022}. One of the challenges is focused on identifying more OCs. 
\cite{Piskunov2006} had estimated that there are about 100,000 OCs in the Milky Way. The number of OCs that have been identified in previous literature is less than one-tenth of the theoretical estimate. A series of studies have searched for OCs using the GAIA 1 and subsequent catalogs. More than 6,000 Galactic star clusters (SCs) have been detected in published Gaia data catalogs \citep{He2022apjs}. 
About 1,200 pre-Gaia open clusters (OCs) have been reidentified \citep{CG2018} based on Gaia Data Release 2~\citep{Gaiadr2}.
A total of 4,000 OCs have been released~\citep{Castro-Ginard2018A, Liu&Pang2019, Castro-Ginard2019, Castro-Ginard2020, Li2022apjs} based on Gaia Data Release 2~\citep{Gaiadr2} and EDR3 \citep{Lindegren2021}.
Recently, \citet{He2022-3} reported 1,656 new star clusters found in the Galactic disk ($\left|b\right| < $ 20 $^{\circ}$) beyond 1.2 kpc, using Gaia EDR3 data.
\cite{Chi2022a} (hereafter Paper II) proposed e-HDBSCAN and reported 83 OCs.

Another challenge is to search for OCs at higher Galactic latitudes. It is rare to find OCs in high-latitude regions of the galaxy, and the majority are found in the thin disk of the galaxy.
Only a few efforts focused on hunting for OCs at higher galactic latitudes and deeper distances. For example, ~\cite{He2022ApJSb} searched all-sky regions nearby ($\varpi > $ 0.8 mas) based on the astrometry of Gaia EDR3 and reported 270 candidates had not been cataloged before, of which 46 clusters are newly found with $\left|b\right| > $ 20 degrees.
\citet{Li2022b} performed a search of high Galactic latitude ($\left|b\right| >$ 20 degrees) with Gaia EDR3 and reported 35 OCs in the high Galactic latitude region with $|b|$ $\geq$ 25$^{\circ}$.
In addition, \cite{sim2019} manually searched the higher-Galactic-latitude regions and identified five clusters at high Galactic latitude with $\left|b\right| >$ 20$^{\circ}$.

It is helpful to search for OCs at higher galactic latitudes since this can provide a better understanding of both OCs and Galactic details outside the Galactic plane. 
There are only a few hundred known OCs at high Galactic latitudes, which is far from the need for statistical studies on properties (i.e., less dust, more distant, and main sequence branch in CMDs) of high Galactic latitude OCs and structural studies of the Milky Way~\citep{Li2022b}.

The release of Gaia data version 3 ~\citep{Gaiadr3} (Gaia DR3) brings us new opportunities to perform star cluster identification. 
Gaia DR3 contains information about object radial velocities (RVs) in the solar system barycentric reference frame ~\citep{Recio-Blanco}. The sample includes 34 million stars with magnitudes in the RV band G$_{RV} \le$ 14. RVs are useful to assess the reliability of the classification of the OC candidate ~\citep{CG2022}. More abundant stellar radial velocity and physical information provide an opportunity to study cluster membership and kinematics~\citep{He2022apjs}. 

In this study, we use a hybrid method (same as Paper I) for blind search that has been proven effective, which combines unsupervised clustering algorithms (FoF and pyUPMASK) and two classification algorithms (RF).
Since the Galactic altitude of OCs can reach $|z|=200-400$ pc ~\citep{CG2018,He2022ApJSb}. 
Additionally, the pursuit of high Galactic latitude OCs  and an examination of their properties contributes to our comprehension not only of OCs, but also of the Galactic regions beyond the Galactic plane. The origin of such OCs remains unclear.
We  extended the search region from the Galactic plane to high Galactic latitudes ($\left|b\right|\geq20$) degrees and attempted a fine-grained blind search based on Gaia DR3 within 5kpc of the solar neighborhood. 

The rest of the paper is structured as follows. Section~\ref{data} describes the data preparation based on Gaia DR3 database. The methodology developed for identifying open clusters is presented in Section ~\ref{method}. Section ~\ref{results} presented all results including 1,179 new star cluster found in the study. The discussions are presented in Section~\ref{discussion}. Finally, a conclusion is presented in Section ~\ref{summary}.

\section{Data Preparation}
\label{data}

We use Gaia DR3 to perform OC blind search in this study. 
Gaia DR3 includes full five-parameter astrometric solutions: positions, parallaxes, and two proper motions for more than 1.468 billion sources (\citet{Lindegren2021}).
In addition, the photometry of the data is available in three photometric bands: G, G$_{BP}$ and G$_{RP}$, which contain sources up to a limiting magnitude of G $\approx$ 21 mags and a bright limit of G $\approx$ 3 mags.

After excluding faint stars ($G< 18$mags) and limiting to parallax ($\varpi$) (from 0.14 kpc to 5 kpc), we obtained more than 20 million stars. In accordance with many previous studies, we selected five parameters (i.e., celestial positions $(l,b)$, parallaxes $(\varpi)$, and proper motions $(\mu_\alpha \cos\delta,\mu_\delta)$) for the identification of open clusters. 
These five parameters of each target source were first normalized. Then the computed results were constructed as a quintet
\begin{equation}
X=\left\{l,b,\varpi,\mu_\alpha \cos\delta,\mu_\delta \right\}
\end{equation}
for each source, respectively.

To better facilitate the clustering calculation and improve the clustering effect, referring to ~\citet{Liu&Pang2019} and Paper II, we constructed a weight for each star, respectively.  

\begin{equation}\label{equ2}
w=\{cos b, 1, 0.5, 1, 1\}
\end{equation}

\begin{equation}\label{equ3}
\begin{aligned}
   X _d &= X \cdot w /(0.2 \cos b+0.7) \\
   &= \frac{\left\{l \cdot \cos b, b,0.5\varpi, \mu_\delta \cos \delta,\mu_\delta\right\}}{0.2 \cos b+0.7} \\
\end{aligned}
\end{equation}
Taking the cosine of $b$ is due to the contraction of $l$ at a given $b$ in spherical geometry, and denominator $(0.2 \cos b+0.7)$ is the normalization factor in the denominator guarantees that $\sum_{i=1}^5 w_i=5$. 
The reason for setting the parallax weight to 0.5 is that the uncertainty of the parallax is greater than the other parameters. Therefore a low weight of parallax can reduce its impact on cluster identification. 

After data preprocessing, we finally obtained a star source dataset $X_d$ with a size of 218,152,787 rows * 5 dimensions for the cluster identifications using the FoF algorithm.

\section{Identification Approaches}
\label{method}

\subsection{Rough Clustering Based on FoF}
Similar to Paper I, we first roughly divided the data into many data regions according to galactic longitude ($l$), galactic latitude ($b$), and parallax ($\varpi$).
The number of divisions for $\varpi$, $b$, and $l$ are 8, 16, and 64, respectively. To avoid splitting the clusters into different regions as much as possible, each of the data regions must not be smaller than two times the typical cluster size  (20 pc) ~\citep{portegies2010young}. To deal with potential clusters located at the boundaries of the region, we set an overlapping region for the two adjacent regions with size ($\varpi$ size 0.2 mas, $l$, and $b$ size 10 pc). 

After applying the above scheme, the whole search volume was divided into 8,596 data regions.
Referring to Paper I, we first performed rough clustering using the FoF algorithm with the linking length ($l_{FoF}$) as 
\begin{center}
\begin{equation}\label{equ4}
     l_{FoF}=b_{FoF} \times\left(\frac{1}{N_{star}}\right)^{1/5} 
\end{equation}
\end{center}
where ${N_{star}}$ is the number of stars in each region, $b_{FoF}$ is the linking length factor. 
According to previous studies~\citep{Liu&Pang2019,Li2022apjs,LISC2,Chi2022}, we set $b_{FoF}$ to 0.2.

Next, we merged the clustering results obtained from each data region.
We adopted a recursive merge strategy to account for clusters at the edge of the data regions. We merge star clusters in two adjacent regions if more than fifty percent of their minimum members are the same.
The overlapping regions in the model were limited to no less than 20 pc (the size of a typical cluster). Therefore, there is typically no intersection point where two clusters share less than 50\% unless the cluster is larger than 60 pc and symmetrically straddles the overlap region established. This is extremely rare.

In merging regions, we set the value of minimum member stars (MMS) to 10, which is inspired by ~\citet{sim2019},~\citet{CG2020}, ~\citet{Hunt2021} and ~\citet{CG2022}.
MMS directly affects the size of the clusters we eventually identify. Therefore, the value determination of MMS is particularly significant. We determined the appropriate MMS from two aspects.
1) According to \citet{sim2019}, the number of members in most cluster are less than 50. 
\citet{Hunt2021} also suggested that the minimum possible size of a star cluster is set to 10 for HDBSCAN, which could detect the majority of OCs in the Gaia data sample.
2) Previous studies revealed that the smallest MMS is 10. In the most widely used open cluster catalog ~\citep{CG2020}, there are almost all (98.5\%(1986/2017)) objects with more than 10 members. Of these, the smallest cluster has only nine members.
In \citet{CG2022}, there are almost all 98.88\% (621/628)) objects with more than ten members.

\subsection{Member Star Determination}
PyUPMASK is a Python package for Unsupervised Photometric Membership Assignment in Stellar
Clusters (UPMASK)~\cite{Krone-Martins2014} used to estimate the membership probability of each input star. 
PyUPMASK has been widely used in the determination of member stars of OC based on astrometric parameters \citep{CG2019, He2022ApJS1, Bai2022RAA, Dias2022}.
The membership probability of stars is iteratively calculated by
\begin{center}
    $P_{star}=\dfrac{K D E_{m}}{\left(K D E_{m}+K D E_{n m}\right)} $,
\end{center}
where $P_{star}$, $KDE_{m}$, and $KDE_{n m}$ are the membership probability of star, KDE of the members,and field star, respectively.
Methods to select reliable cluster members by membership probability have been widely used, such as ~\cite{Jaehnig2021}, ~\cite{He2022apjs,He2022ApJSb,He2022ApJS1},~\cite{Niu2020ApJ}  etc.
Cut stars with membership probability p $<$ 0.5, which is an optimal threshold ~\citep{Soubiran2018} and~\citep{Carrera2019}, to reduce the contamination of field stars.
Analogously, after  feeding $X_d$ to the FoF clustering model for get many rough cluster candidates, we have a membership census in each candidate with pyUPMASK and keep members whose membership probability is greater than 0.5 which is same as ~\citet{Jaehnig2021} but smaller than that in ~\citet{Gao2018AJ}.

\subsection{Open Cluster Identification}
We performed a series of processing to analyze and identify OCs.
We first selected OC candidates by RF model; we then  filtered the results using proper motion dispersion;  after excluding the published OCs by cross-matching, we made an isochrone fitting and classified the OCs; we finally obtained credible OCs using manual inspection.

\subsubsection{Random forest modelling}
We used random forest (RF) model to isolate the most likely cluster members based on membership probabilities without normalizing high-dimensional data.
Same to Paper I, in the third step, we trained a Random Forest model with samples collected in the work of \cite{Cantat-Gaudin2020} with Gaia DR2 and Gaia EDR3 \citep{Castro-Ginard2018A, Castro-Ginard2019, Castro-Ginard2020, Castro-Ginard2022} to detect OCs among the potential candidates.

\subsubsection{Filtering using proper-motion dispertion}
To select potentially real OCs from spatial over-density structures, we used the following proper-motion criterion \citep{Hunt2021,Hao2022,Hao2022b} and Paper I.
\begin{equation}\label{equ5}
     \sqrt{\sigma_{\mu_{a^{*}}}^{2}+\sigma_{\mu_{\delta}}^{2}} \leq 0.5 \mathrm{mas}\  \mathrm{yr}^{-1} \text { if } \varpi<1 \text { mas }  
\end{equation}
\begin{equation}\label{equ6}
     \sqrt{\sigma_{\mu_{a^{*}}}^{2}+\sigma_{\mu_{\delta}}^{2}}  \leq 2 \sqrt{2} \frac{\varpi}{4.7404} \mathrm{mas} \  \mathrm{yr}^{-1} \text {   if } \varpi \geq 1 \text { mas }
\end{equation}
where $\sigma_{\mu_{a^{*}}}^{2}$ and 
$\sigma_{\mu_{\delta}}^{2}$ are the dispersion in positional space $\mu_{a^{*}}$ and ${\mu_{\delta}}$, respectively.
It should be noted that this criterion is necessary for the identification of OCs. It does not imply that clusters that have dispersion higher than the criterion in these formulas are true clusters. However, by doing so, we can filter out most of the candidates that do not meet this criterion, obtain high quality cluster candidates, and reduce the final identification effort.

\subsubsection{Cross-match}
To exclude as many reported clusters as possible, we cross-matched them with pre-Gaia cluster catalogs, OC catalogs based on Gaia, and globular cluster catalogs.
The pre-Gaia cluster catalogs (MWSC) contained 3006-star clusters gathered by \cite{Dias2012} and \cite{Kharchenko2013},  aggregated from various data sources. 
Since there is no relevant proper motion parameter in the data, we have to only compare the mean parameters within 5$\sigma$ (where$\sigma$ is the uncertainty listed in both catalogs for each quantity) using sky coordinates.
To eliminate as many OCs as possible that have already been found and obtain OC candidates that have not been unnoticed before, we consider an OC to be positionally matched to a cataloged one if their astrometric mean parameters (${l,b,\varpi,\mu_\alpha,\mu_\delta }$) are compatible within 5 $\sigma$ (where $\sigma$ is the uncertainty quoted in both catalogs for each quantity) which is consistent with \cite{He2022apjs}, \cite{He2022ApJSb} and \cite{Hao2022b}.
The list of previously published sources including LP, Ferreira Series, CWNU, Hao Series, UBC, and so on is presented in Table~\ref{tab:list_OCs}.

We also carried out the same method to cross-match with Globular clusters (GCs) to exclude globular clusters.

\subsubsection{Isochrone fitting and classification}
\label{iso-fit}
We performed isochrone fitting for each new result, following the methods described in Paper I. The PARSEC theoretical isochrone models~\citep{Bressan2012} have been updated by the Gaia EDR3 passbands using the photometric calibrations from ESA/Gaia. The extinction curve of $\mathrm{R}_{\mathrm{V}}=3.1$ has been reddened to derive their physical parameters (age and metallicity). 

A log-normal initial mass function~\citep{Chabrier2003} is used to generate an isochrone library from $log(\dfrac{t}{yr})=$ 6.0 to 11.13 at steps of $\Delta(log t ) = 0.03 $ while metal fractions from 0.002 to 0.042 with a step of 0.002.
An objective fitting function
\begin{center}
\begin{equation}
    \bar{d}^{2}=\dfrac{\sum_{k=1}^{n}\left(\mathbf{x}_{k}-\mathbf{x}_{k, n n}\right)^{2}}{n}
\end{equation}
\end{center}
was applied to all new OC candidates, where $n$ is the
number of selected members in a cluster candidate, and ${x}_{k}$ and
${x}_{k, n n}$ are the positions of the member stars and the points on the
isochrone that is closest to the member stars, respectively.
A series of isochrones are produced with the  parameters in Table \ref{parameters_set} referred to~\citet{Chi2022}.
\begin{table}[H]

\centering
\caption{The range and step of Age and Z for Isochrone Fitting}
\begin{tabular}{lccr}
\toprule
    Parameter & Range& Step & Unit \\ 
\midrule
Age    &   6.0 -- 10.11  & 0.03 & log(t/yr)\\
Z      &  0.002 -- 0.042       & 0.002& $[\mathrm{Fe} / \mathrm{H}]$      \\

\bottomrule
\end{tabular}
\label{parameters_set}
\end{table}
After the isochrone fitting, we classified the fit results to facilitate the identification of well-fitting OCs.
We performed this by calculating the dispersion $\sigma_{d^2}$ of $d^2$ by
\begin{center}
\begin{equation}
\sigma_{d^2}=\sqrt{\frac{\sum_{k=1}^n\left(d_k^2-\bar{d}^2\right)^2}{n}}
\end{equation}
\end{center}
and $\sigma_{d^2}$ reflects how close the core sample is along the isochrone.

We subsquently  classified them in to 3 categories (class A, class B and class C) based on isochrone-fitting according to stringent criteria listed in Equation~\ref{classify_label}.
\begin{center}
\begin{equation}
\label{classify_label}
\text { Class }= \begin{cases}A & \left(r_n<0.1\text { and } \bar{d}^2<0.02 \text { and } \sigma_{d^2}<0.04\right) \\ B & \left(r_n<0.1  \text { and } \bar{d}^2 \geqslant 0.02 ) \text { or }(r_n<0.1 \text { and }\sigma_{d^2} \geqslant 0.04\right) \\ C & \text { others }\end{cases}
\end{equation}
\end{center}

Ones with clear CMDs ($r_{\text{n }}<0.1$, $\sigma_{d^2}<0.04$ and $\bar{d}^2<0.02$) and enough bright star members, which have more than 20 members of magnitude less than 17, are class A.
Same as class A but with unclear isochrone ( $r_{\text{n }}<0.1$ ) are class B.
The rest of the candidates with a loose CDM distribution are class C.
One can refer ~\citet{Liu&Pang2019} and Paper I for more details about the calculation of $r_{\text{n }}$. 

\subsubsection{Comprehensive Analysis Based on Visual Insection}
Since a real OC should have clear main sequence features on the CMDs, reference \citet{CG2020} and \citet{He2022apjs}, to screen out the most reliable candidates, we performed manual visual inspections on spatial distributions (SDs), proper motion distributions (PMDs), parallax distributions (PDs) and $\varpi$ vs $\mu_\alpha$ , and their isochrone fits results to further check the quality of candidate clusters.

\section{Results} 
\label{results}

In data sources of 218,152,787 stars generated by pre-processing, we obtained 14,701 stellar aggregates.
23 aggregates were rejected by the RF model, and 1,063 aggregates were eliminated by proper-motion dispersion filtering.
Among them, 1,244 OCs can be cross-matched with published catalogs, and there are 12,371 likely new OCs that need further identification (i.e., isochrone-fitting and visual inspection), of which 12,316 are located at latitudes of $\left |b\right|<20$ degrees, and 55 OCs are located at latitudes of $\left|b\right|>20$ degrees.

We further carried out an isochrone-fitting and divided all possible clusters (12,371 OCs) into three classes based on the approach mentioned in subsection~\ref{iso-fit}, i.e., 1,194 (class A, see Figure~\ref{fig:classA}), 5,252 (class B), and 5,925 (Class C). Class A means the OCs have clear main sequences in the CMD and more star members, which is also what we focus on. 

After manual visual inspection, 1,179 OCs are supposed to be real galactic star clusters.
Derived astrophysical parameters for the final OCs are given in Table ~\ref{tab:tab1}.
All member sources are presented in Table~\ref{tab:sources}.

\begin{figure}[htbp]
    \subfigure{
    \includegraphics[width=7in,height=1.7in]{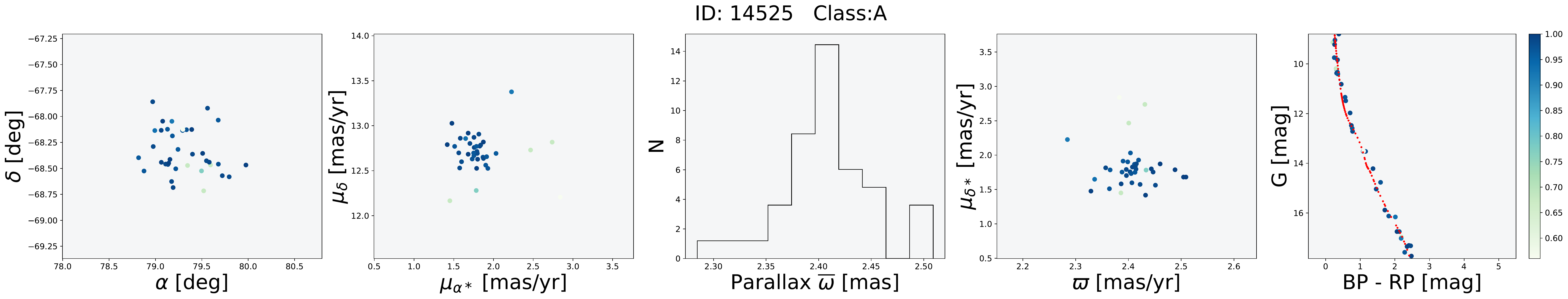}
    }
    
    \caption{An example of high Galactic latitude OCs.}
    \label{fig:hb}
\end{figure}

In addition, as far as we know, only a few star clusters have been discovered at high galactic latitudes of $\left| b\right| \geq 20$ degrees.
Based on manual visual inspection, ID3252 and ID14525 is identified as a true high galactic latitude OC.
Those 5-panel plots are presented in Figure~\ref{fig:hb}.
The remaining 53 clusters which are located at high galactic latitudes are not listed in this work because they do not have clear main sequence features.

\section{Discussions}
\label{discussion}

\subsection{Plausibility analysis}
\begin{figure}[htbp]
\centering
    \subfigure[]{
    \includegraphics[width=2.6in,height=2.2in]{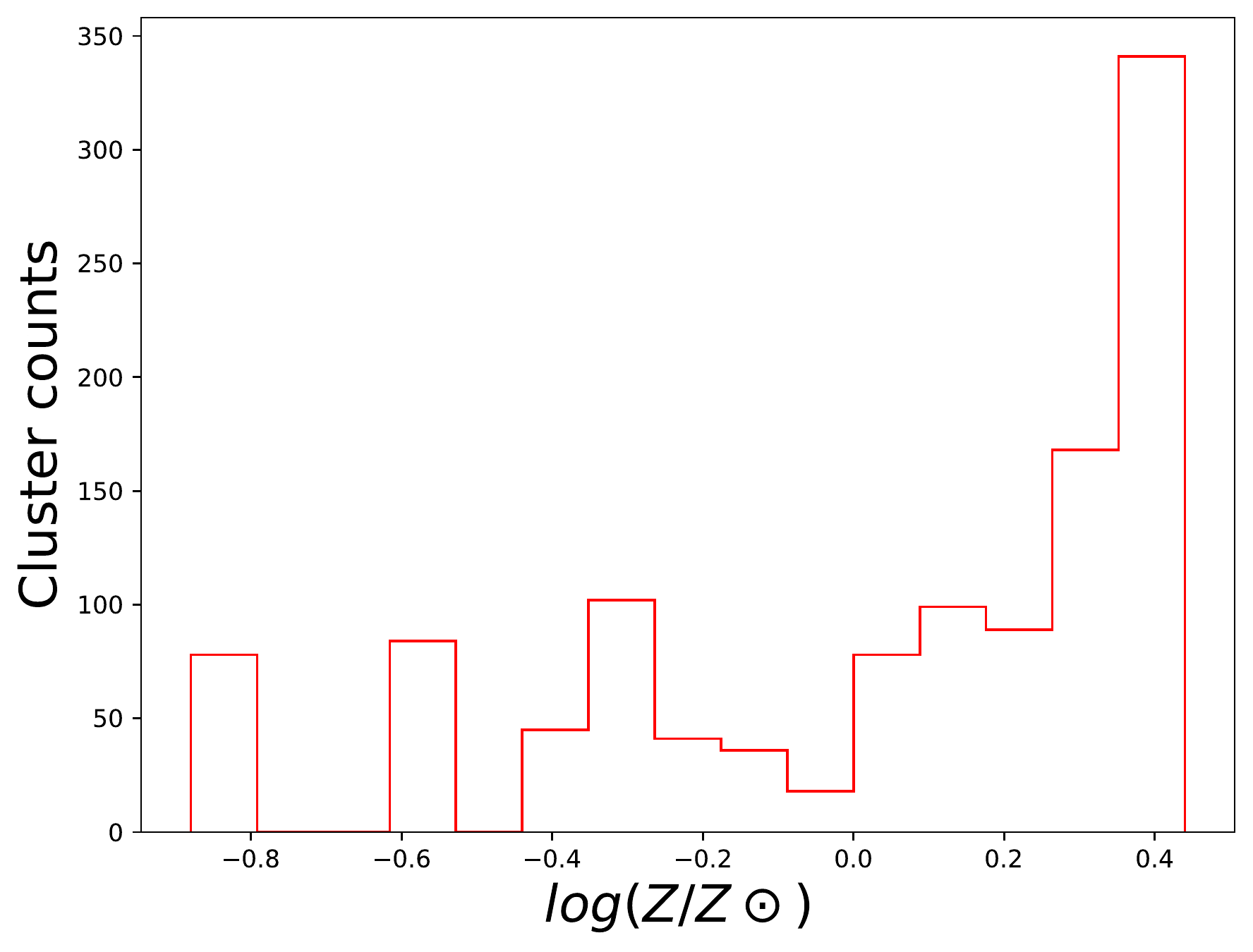}
    }
    \subfigure[]{
    \includegraphics[width=2.6in,height=2.2in]{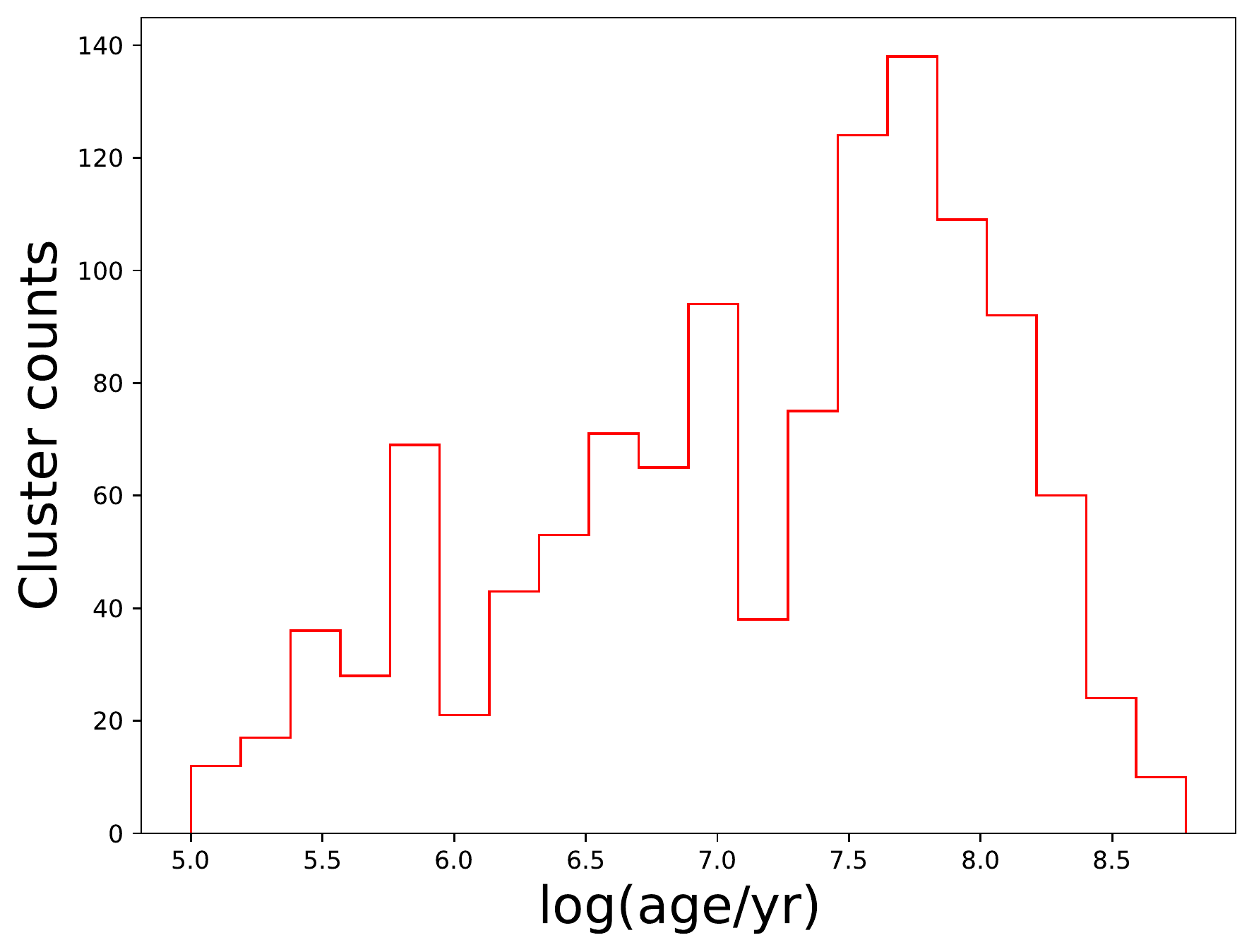}
    }
    \subfigure[]{
    \includegraphics[width=2.6in,height=2.2in]{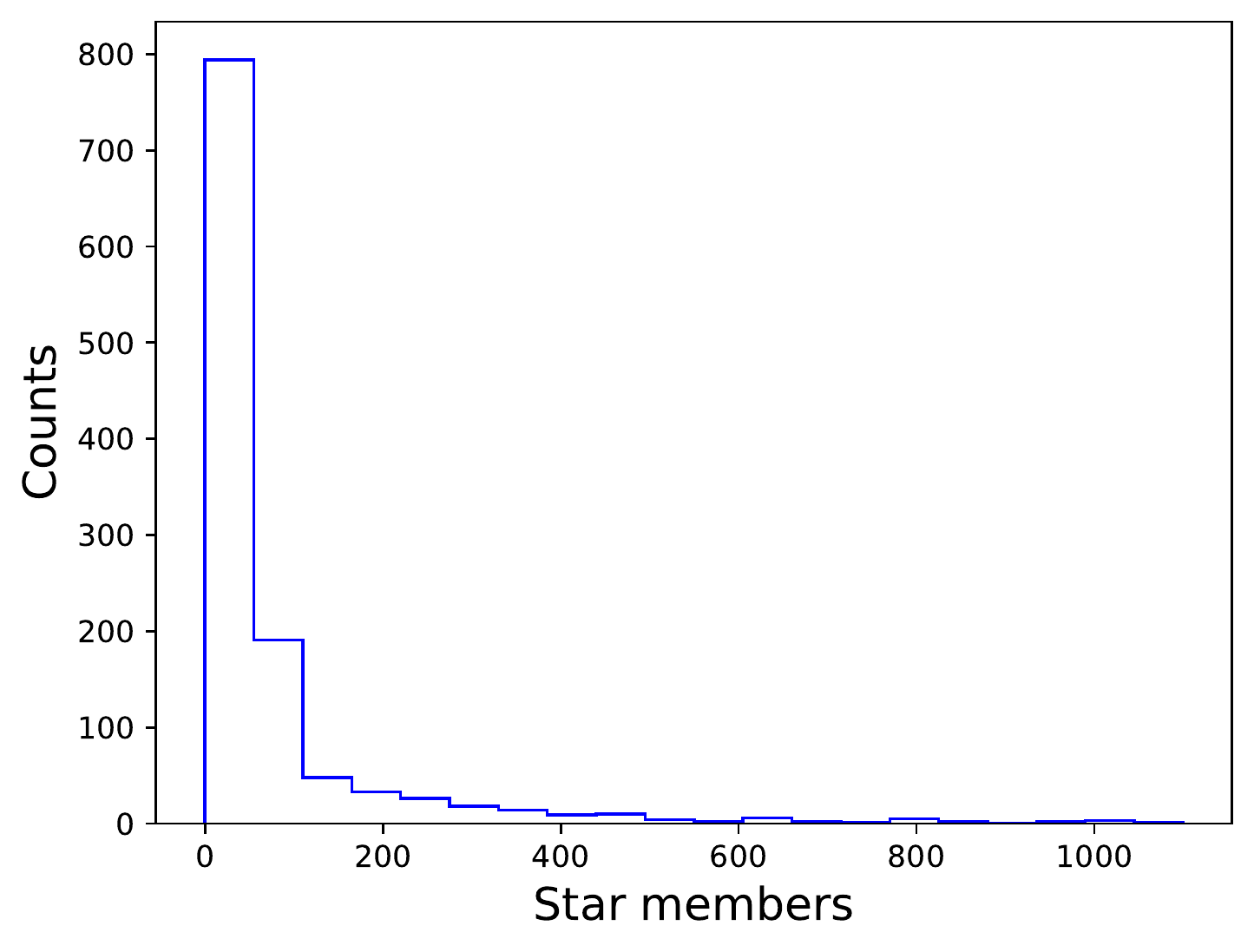}
    }
    \subfigure[]{\includegraphics[width=2.6in,height=2.2in]{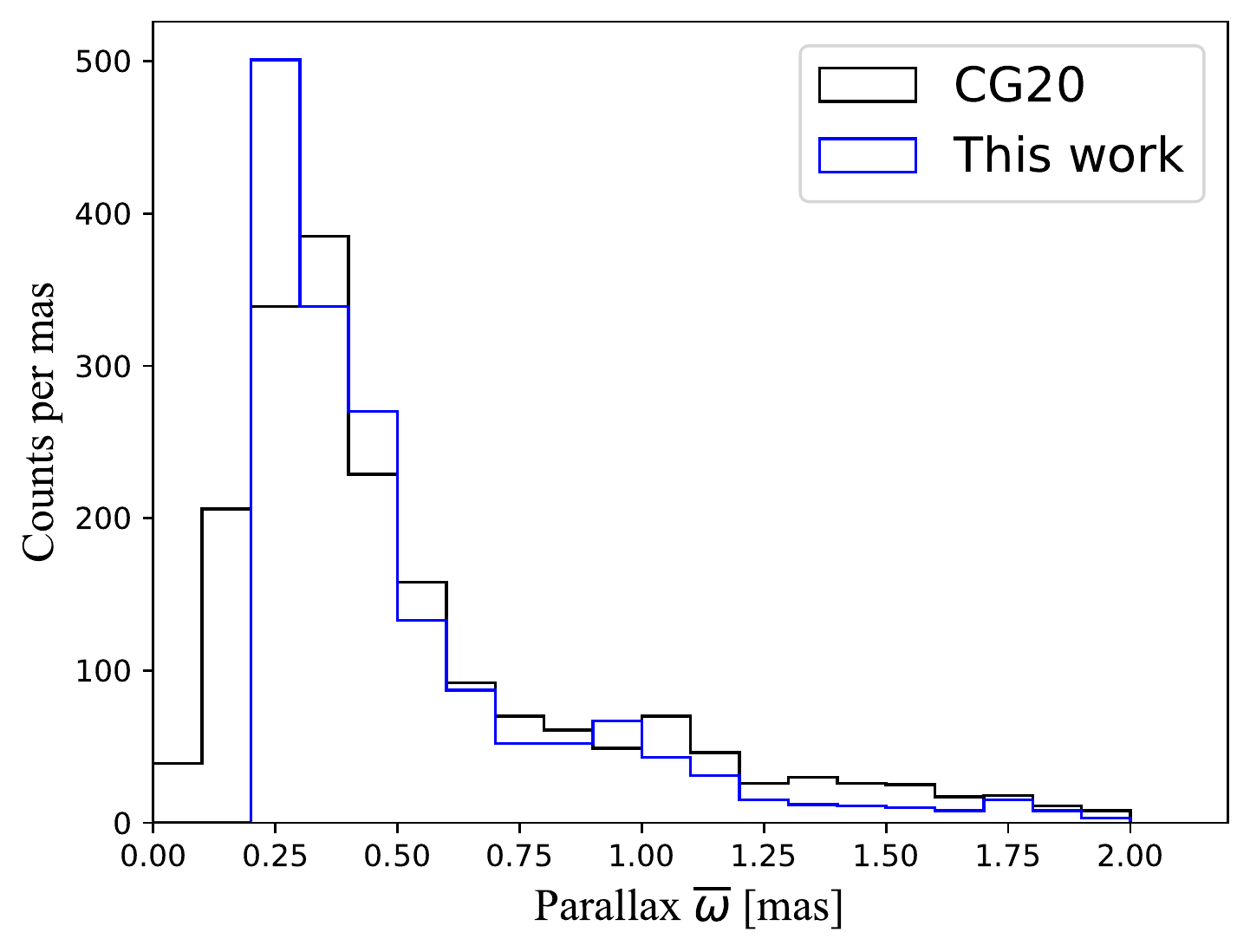}}
    
    \caption{The histogram of Z and the age of 1,179 OCs were showed in (a) and (b), respectively. (c) and (d) present the member and parallax distribution of 1,179 OCs, respectively.}
    \label{fig:Z_age}
\end{figure}

The results of the 1,179 OCs we found are reasonably plausible.
The identified new reliable 1,179 cluster candidates in Class A account for 8.02\% of all possible clusters in this work. 
This ratio is similar to 3.11\% (76/2443) of \citet{Liu&Pang2019}, which used FoF to identify star clusters in Gaia DR2.
On the other hand, compared with ~\citet{Liu&Pang2019}, the fitting criteria we adopted were more stringent, for example, the $\bar{d}^2$ value was more strictly restricted (~\citet{Liu&Pang2019} is 0.04, ours was 0.02). As a result, thanks to  the huge volume of high-quality data of Gaia DR3, the number of clusters that are eventually identified is more, but the quality and reliability are higher.
As shown in  Figure~\ref{fig:Z_age} (c),  most of the OCs member stars we report are in the intermediate distribution of less than 100, with 740 clusters with less than 50 members, (63 percent of the total). 
The number of clusters with fewer than 30 members is 228, which is 20 percent of the total.
This indicates that our fine-grained blind search method can effectively detect small OCs.

\begin{figure}[htbp]

    \centering
    \subfigure[]{\includegraphics[width=2.6in,height=2.2in]{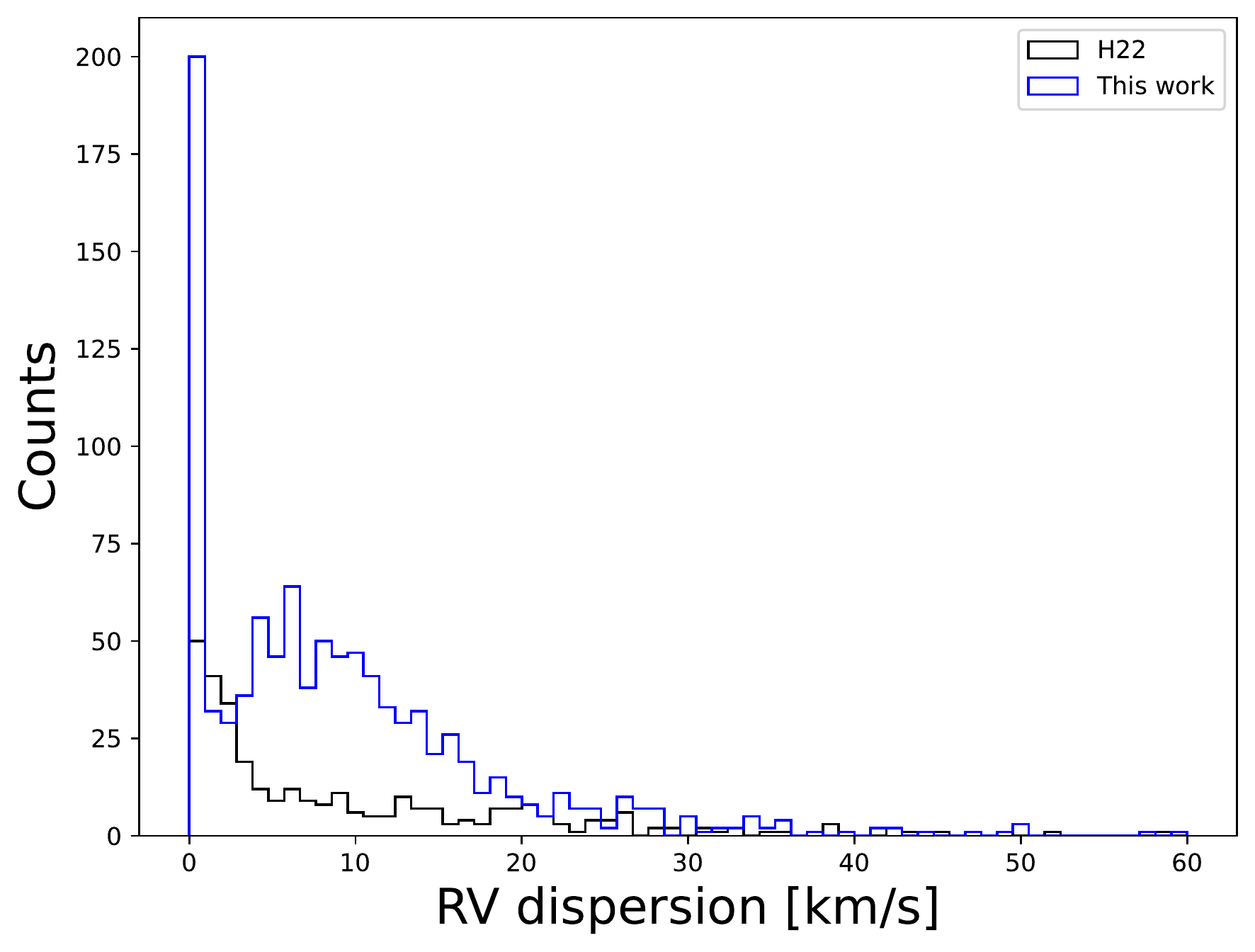}}
     \subfigure[]{\includegraphics[width=2.6in,height=2.2in]{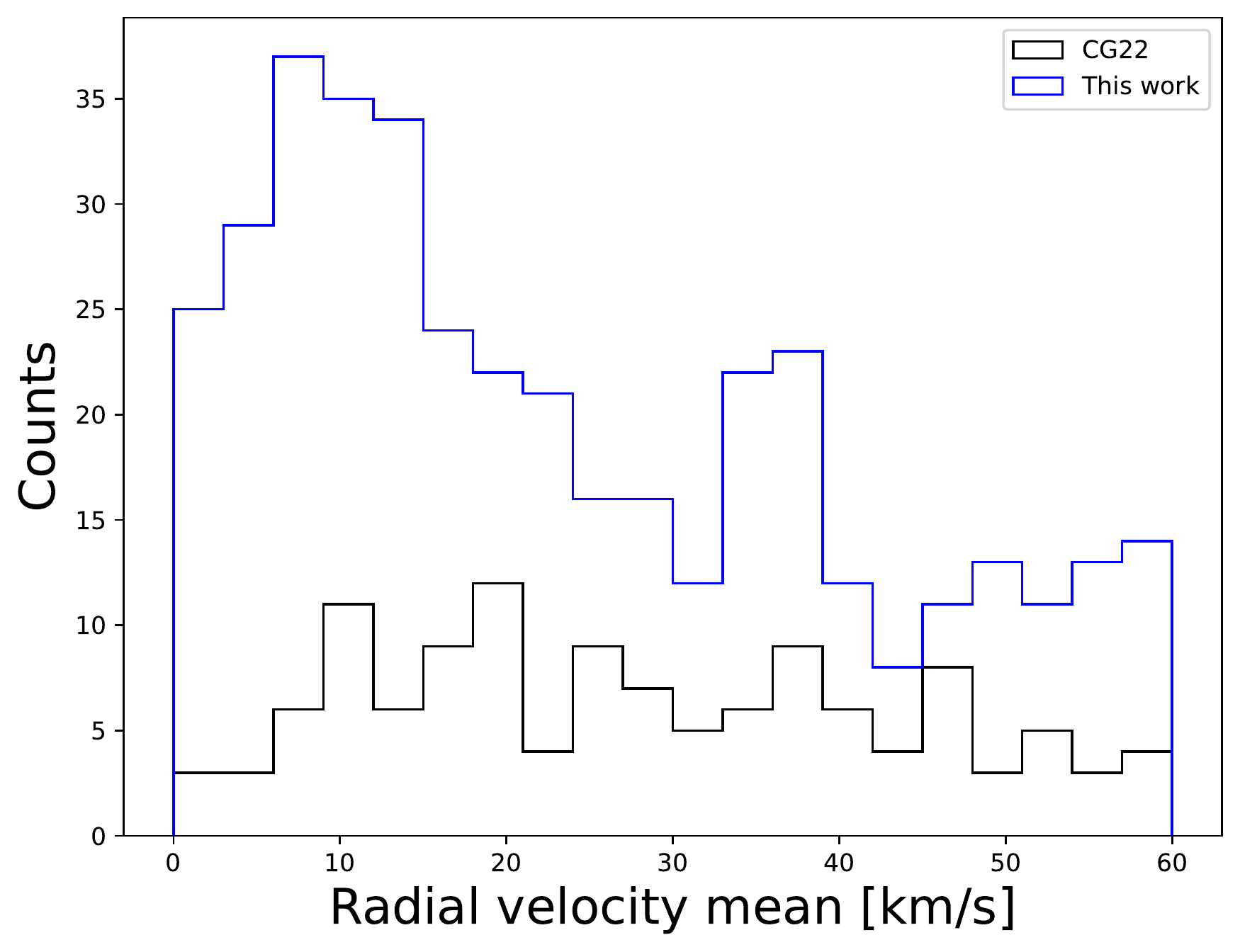}}
     \subfigure[]{\includegraphics[width=2.6in,height=2.2in]{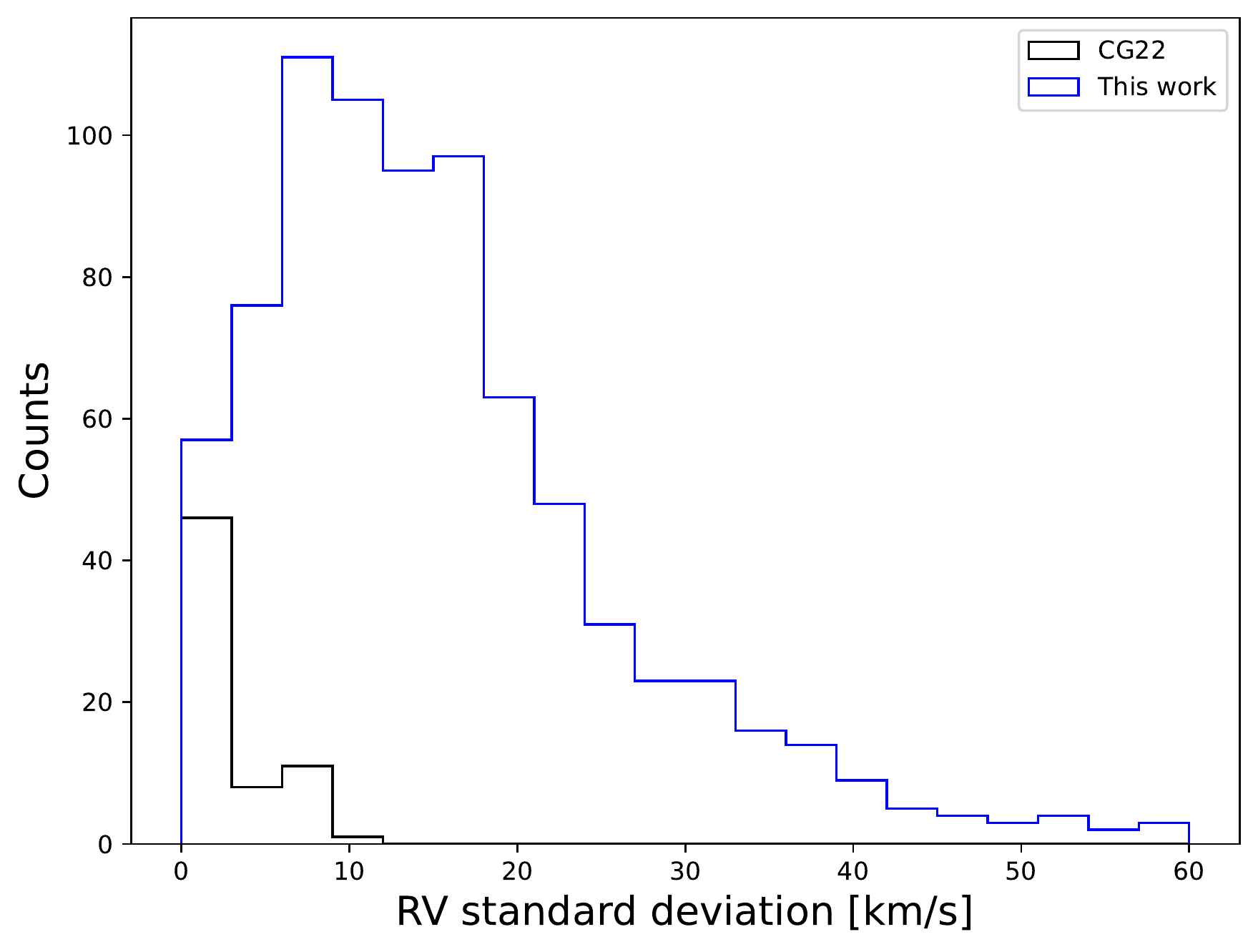}}
    \subfigure[]{\includegraphics[width=2.6in,height=2.2in]{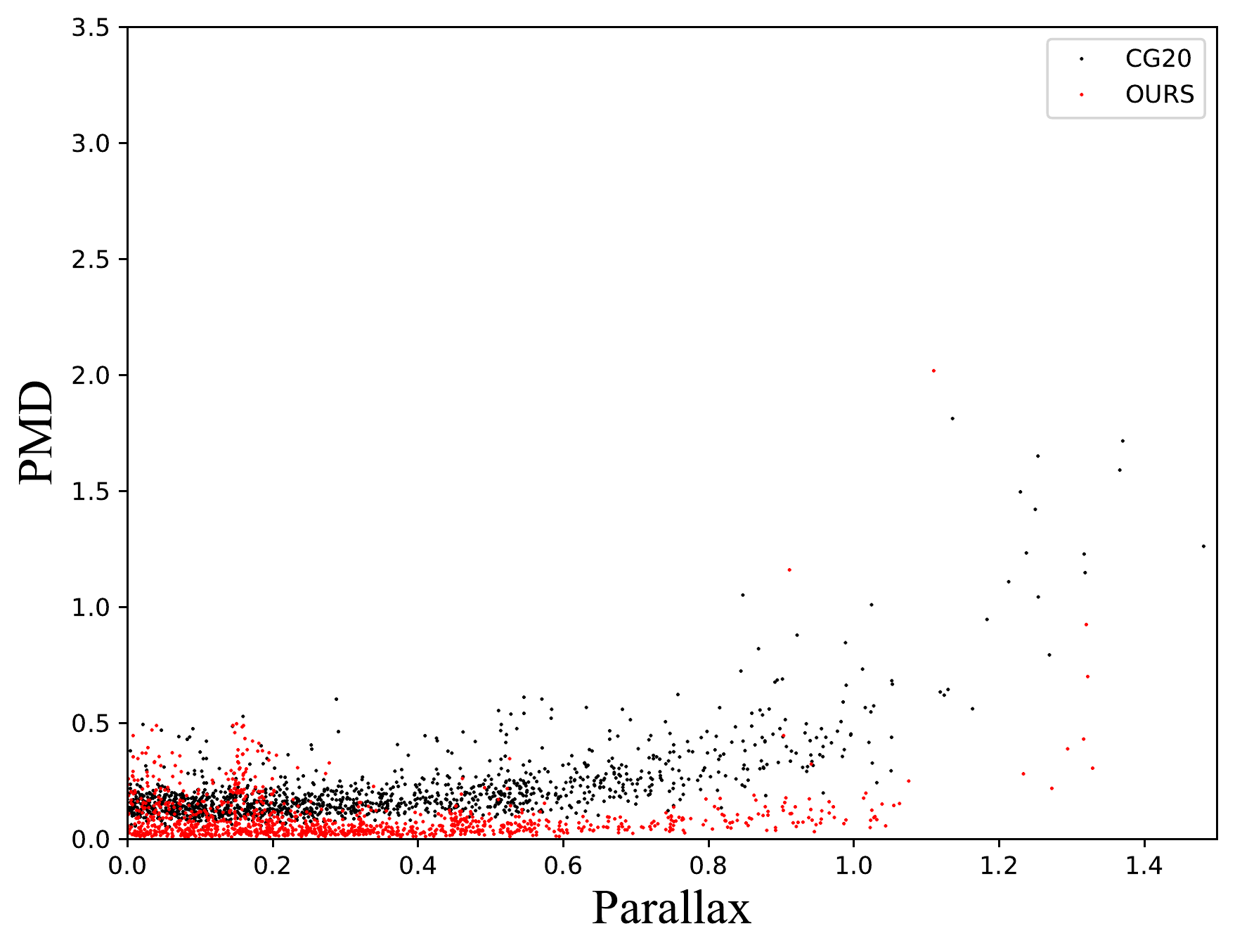}}
    \caption{RV dispersion distribution diagram (a) and mean distribution diagram (b) of 1,179 OCs compared with H22. RV standard deviation distribution diagram (c) and PMD distribution diagram (d) of 1,179 OCs compared with CG22.}
    \label{fig:rv_dispersion_distritution}
\end{figure}

To validate our results found in this study, we compared them to the OCs in CG20 with parallax distribution (see  Figure~\ref{fig:Z_age} (d)).
Compared with CG20, our clusters are mostly above 0.25 and  similar in their number, size, and distribution.
The peak of the median distribution of our results (990) (see  Figure~\ref{fig:rv_dispersion_distritution} (a)) is consistent with H22, which is local at 1 km s$^{-1}$. Most parts of RV dispersion are smaller than 13 km s$^{-1}$.
Right subgraph of Figure~\ref{fig:rv_dispersion_distritution} (b) shows the  mean RV distribution of our 1,179 OCs is consistent with ~\citet{CG2022} (CG22) which has 170 OCs with RV. The mean RV of our 1,179 OCs is 4.477 km s$^{-1}$, while CG22 is 31.038 km s$^{-1}$.
Left subgraph of Figure~\ref{fig:rv_dispersion_distritution} (c) shows the distribution of RV of our 1,179 OC (809 OC have more than 2 member stars with RV data). The standard deviation is consistent with CG22 (66 OCs have more than 2 member stars with RV data).
As \cite{CG2020}, \cite{1743oc} and \cite{He2022ApJSb} stated, proper motion dispersions (PMD) of cluster members are a measure of whether a candidate SC is a true cluster or not.
From right subgraph of Figure~\ref{fig:rv_dispersion_distritution} (d), the proper motion dispersions of our results are well consistent with but smaller than the value of ~\citet{CG2020}.
Figure~\ref{fig:Z_age} (b) shows the age and Z distributions of the SC candidates. It is obvious that the new SC candidates are younger than 8.5 log(age/yr). 
And the left panel (Figure~\ref{fig:Z_age} (a)) shows that many of the OCs are metal-poor (smaller than 0.4 $log (Z/ Z\odot)$).

From the comparison of the above four aspects, combined with the CMD properties of the original candidates, it is clear that the current candidates are clusters with the characteristics of genuine SCs.

    

\subsection{Member Size Analysis and Comparison}
\label{sec:matched_ocs}
\begin{figure*}[htbp]
    \begin{center}
    \subfigure{
    \includegraphics[width=7in,height=1.7in]{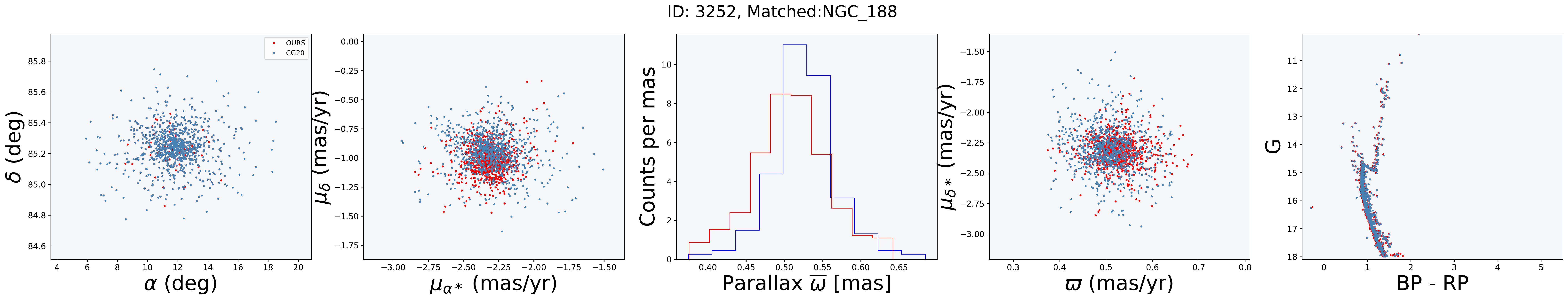}
    }
    \subfigure{
    \includegraphics[width=7in,height=1.7in]{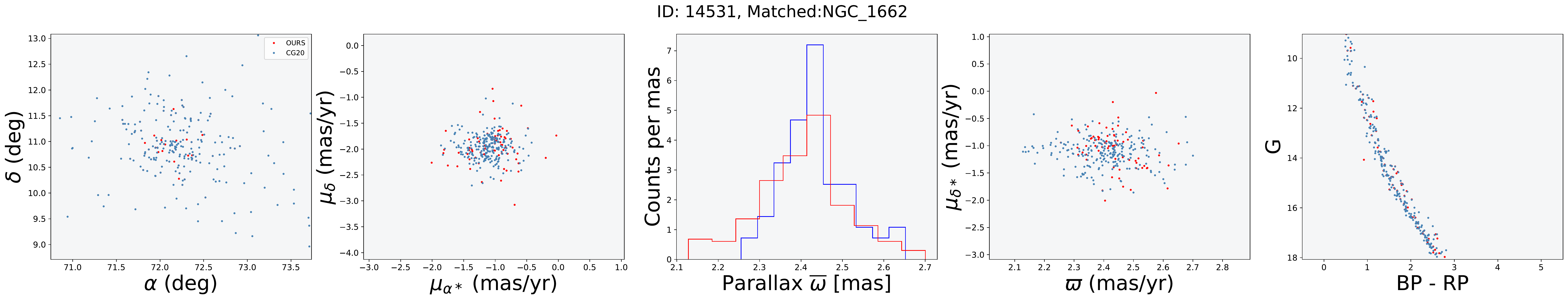}
    }
    \subfigure{
    \includegraphics[width=7in,height=1.7in]{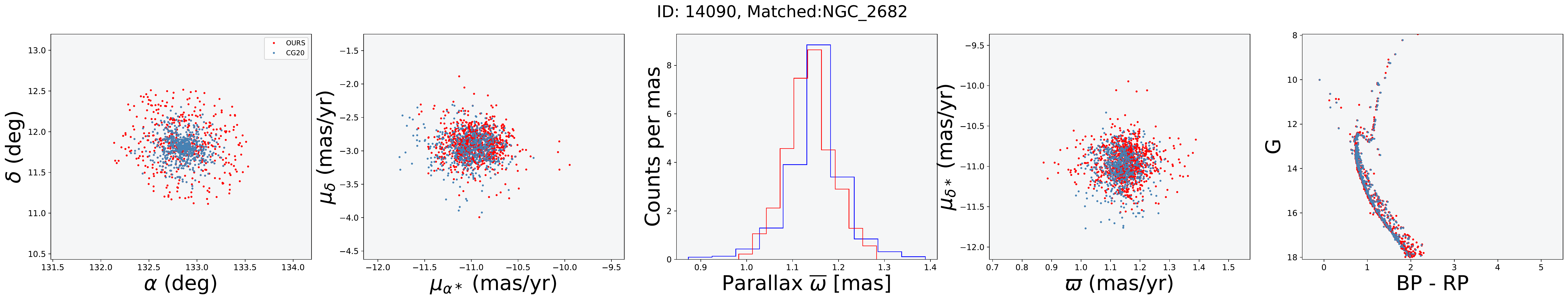}
    }
    \subfigure{
    \includegraphics[width=7in,height=1.7in]{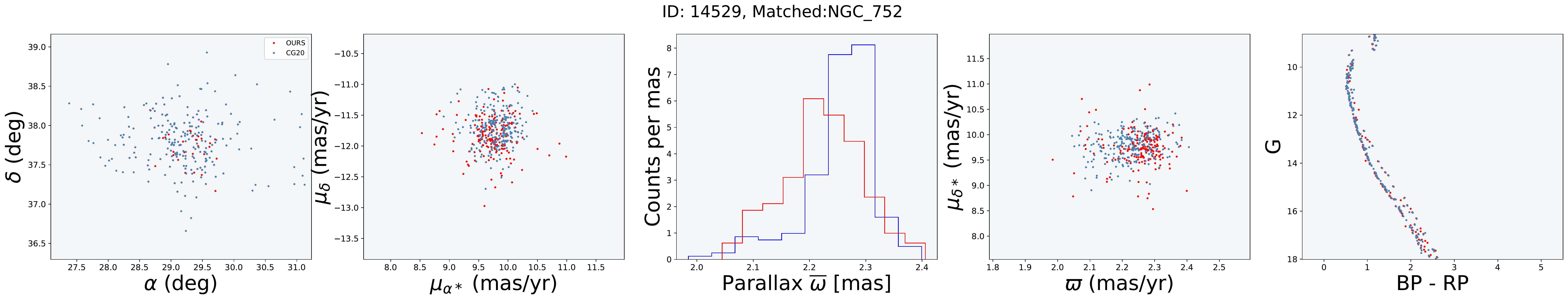}
    }
    
    \end{center}
    \caption{Schematic diagram of the results (i.e., NGC188, NGC1662, NGC2682, and NGC752) of matching with CG20.}
    \label{matched_ocs}
\end{figure*}

For most of the matched clusters, we found that the clusters have more member stars than those identified in Gaia DR2.
Figure~\ref{matched_ocs} (a) and (d) indicate that there is more concentrated membership around the cluster halo. We can re-detect more members in NGC2682 (M67), which are located in more concentrated areas and less contaminated by field stars (see Figure~\ref{matched_ocs} (c)). 
We also discovered more members further away from the center in NGC1662 while maintaining good CMD main sequence characteristics (see Figure~\ref{matched_ocs} (b)).
That means previously reported cluster scales may be underestimated, which is consistent with the work of ~\cite{Zhong2022}.

\subsection{Deficiencies}
\begin{figure}[htbp]
    \centering
    \includegraphics[width=6.0in,]{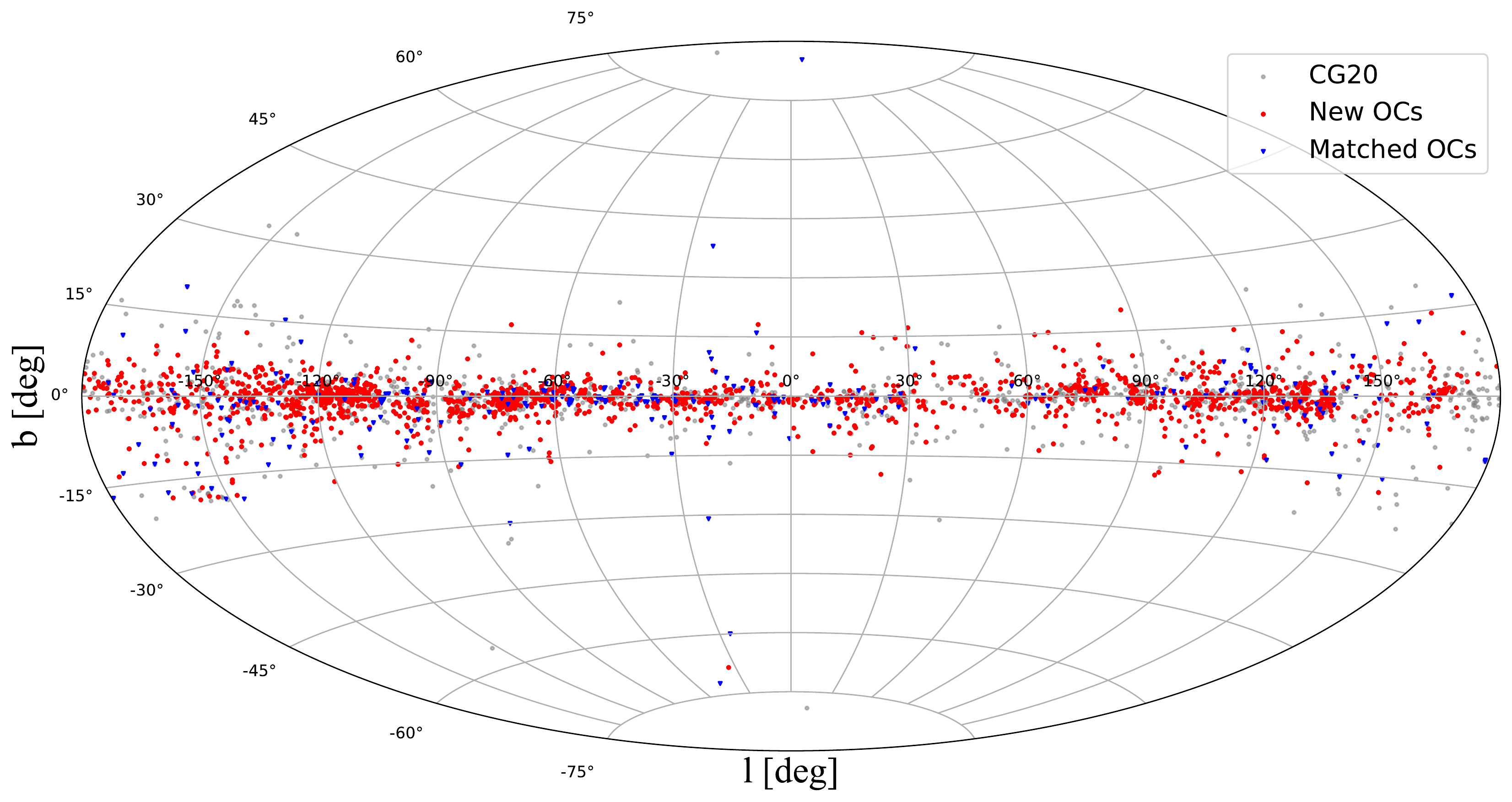}
    \caption{Distribution of the OC candidates. Matched known clusters (blue triangles) , the CG20 OCs (gray dots) and the new 1,179 OCs are given respectively.}
    \label{fig:oc_distribution}
\end{figure}

\begin{figure*}[htbp]
    \begin{center}
    \subfigure{
    \includegraphics[width=7in,height=1.5in]{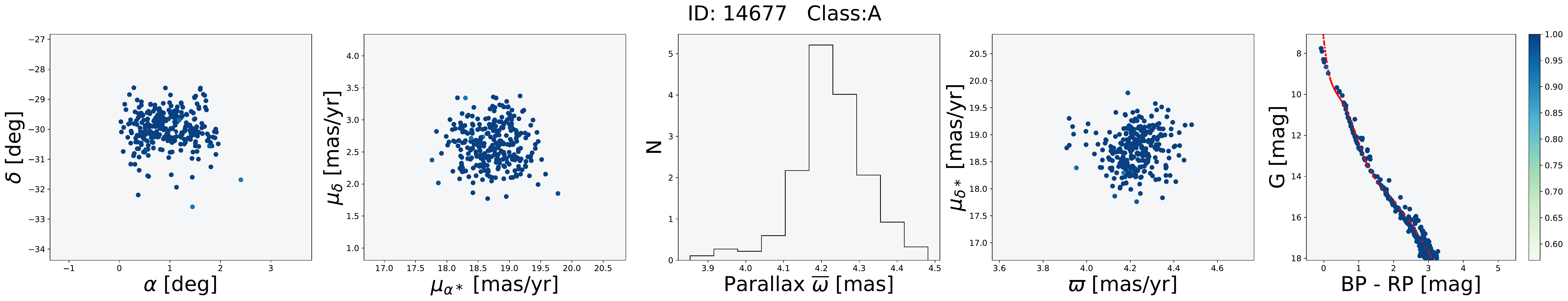}
    }
    \subfigure{
    \includegraphics[width=7in,height=1.5in]{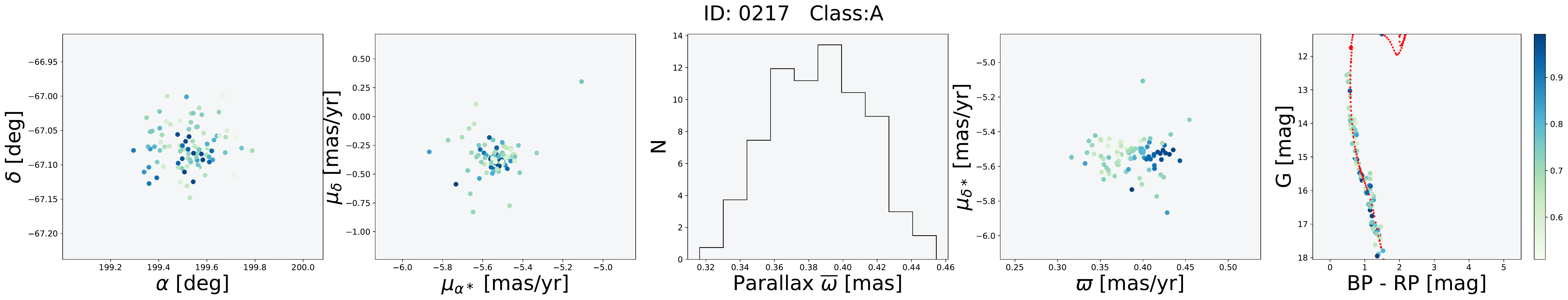}
    }
    \subfigure{
    \includegraphics[width=7in,height=1.5in]{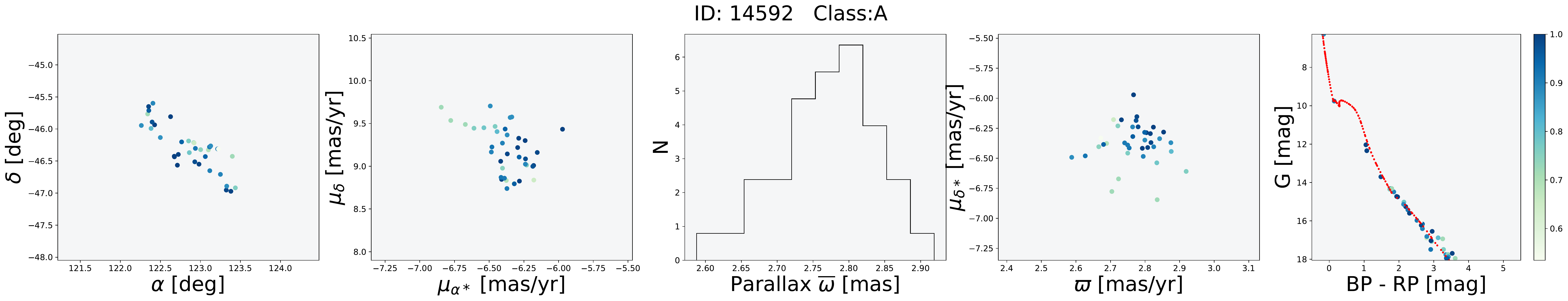}
    }
    \subfigure{
    \includegraphics[width=7in,height=1.5in]{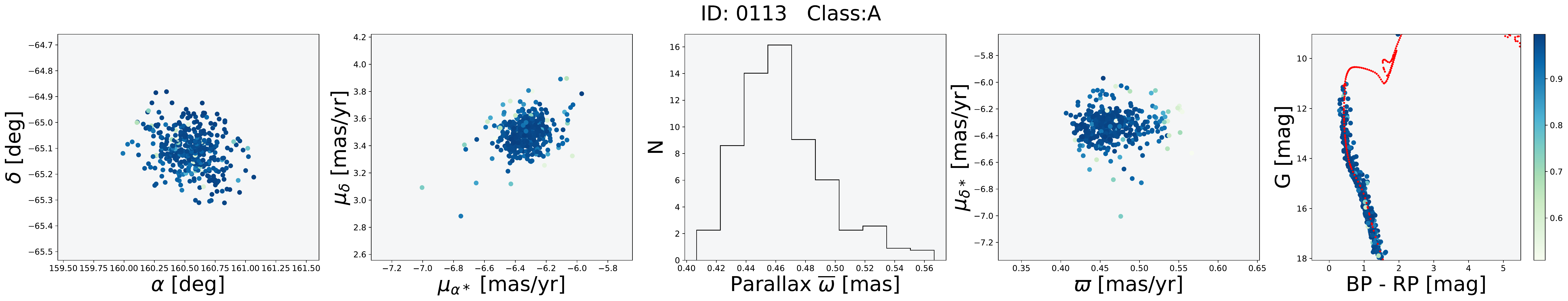}
    }
    \subfigure{
    \includegraphics[width=7in,height=1.5in]{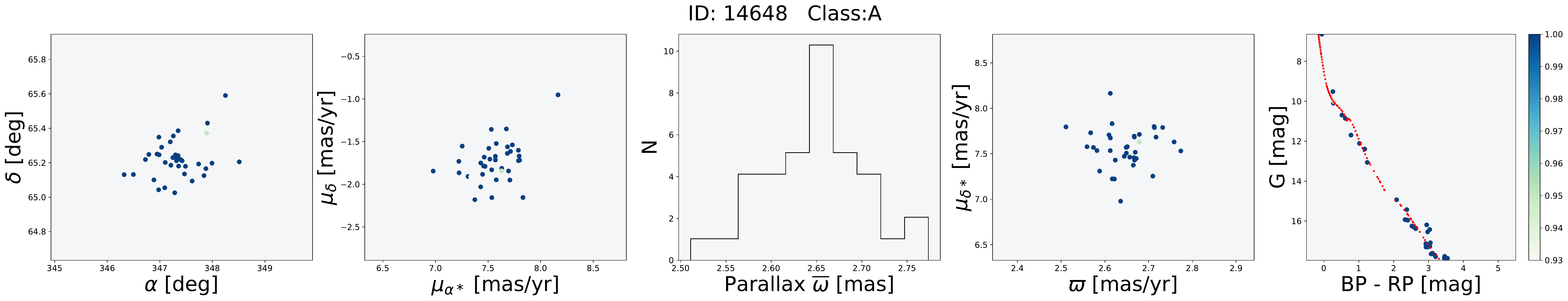}
    }
    \end{center}
    \caption{Examples of class A. From left to right, the subplots show the spatial distribution, proper-motion distribution, parallax statistics, parallax distribution and CMD with best-fitting isochrone line, respectively.
    The color bars represent the cluster probability  of the member stars. }
    \label{fig:classA}
\end{figure*}

\begin{figure*}[htbp]
    \begin{center}
    \subfigure{
    \includegraphics[width=7in,height=1.5in]{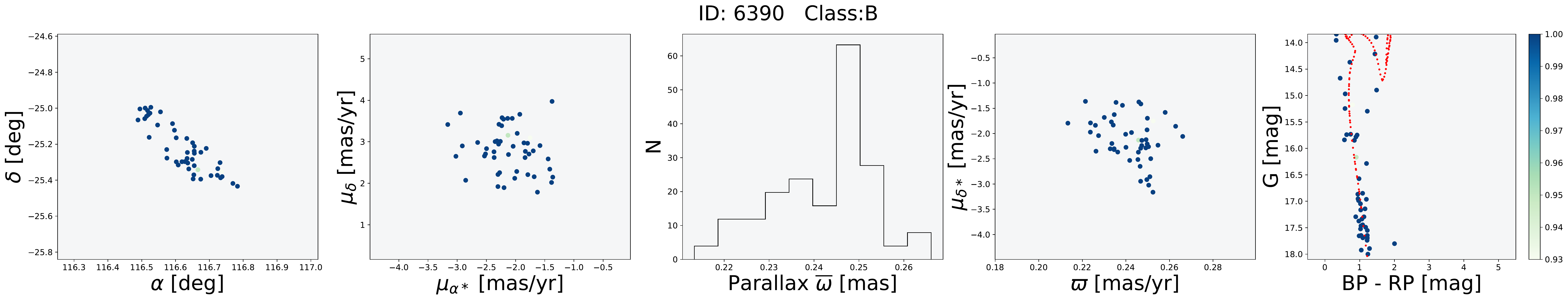}
    }
    \subfigure{
    \includegraphics[width=7in,height=1.5in]{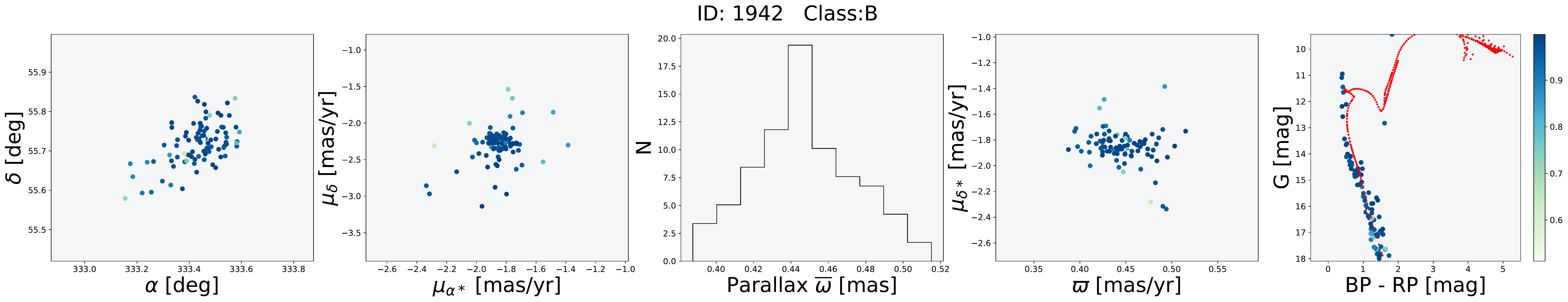}
    }
    \subfigure{
    \includegraphics[width=7in,height=1.5in]{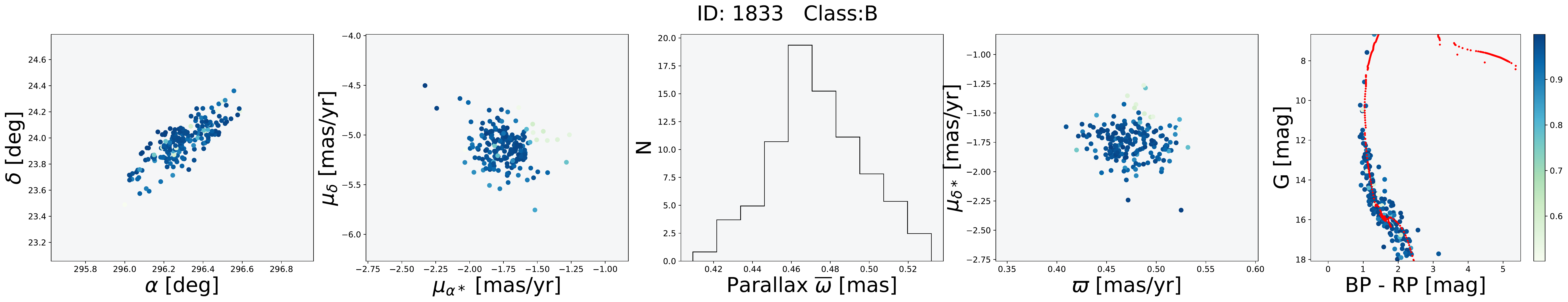}
    }
    \subfigure{
    \includegraphics[width=7in,height=1.5in]{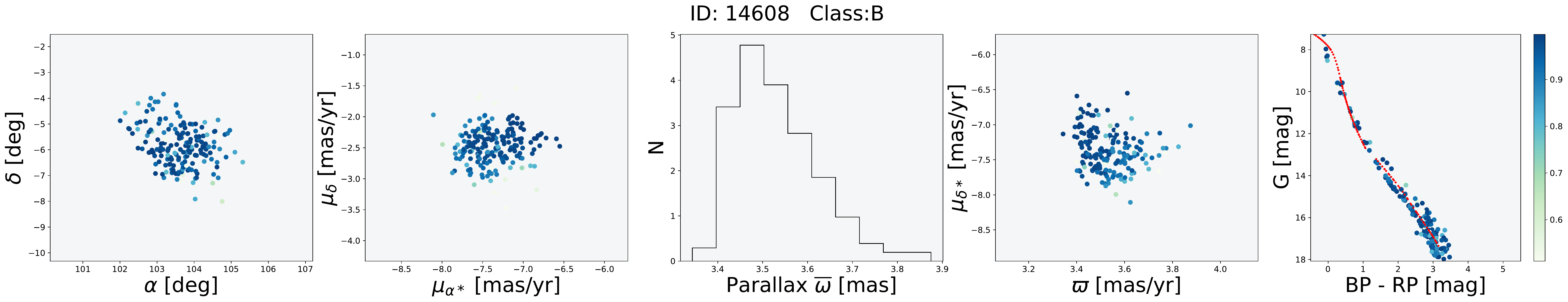}
    }
    \subfigure{
    \includegraphics[width=7in,height=1.5in]{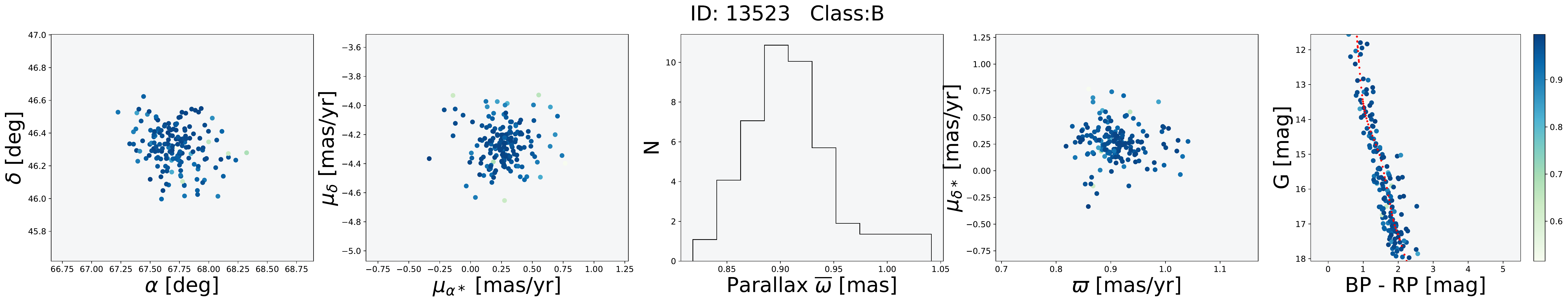}
    }
    \end{center}
    \caption{As same as Figure~\ref{fig:classA}, but for Examples of class B. }
    \label{fig:classB}
\end{figure*}

\begin{figure*}[htbp]
    \begin{center}
    \subfigure{
    \includegraphics[width=7in,height=1.5in]{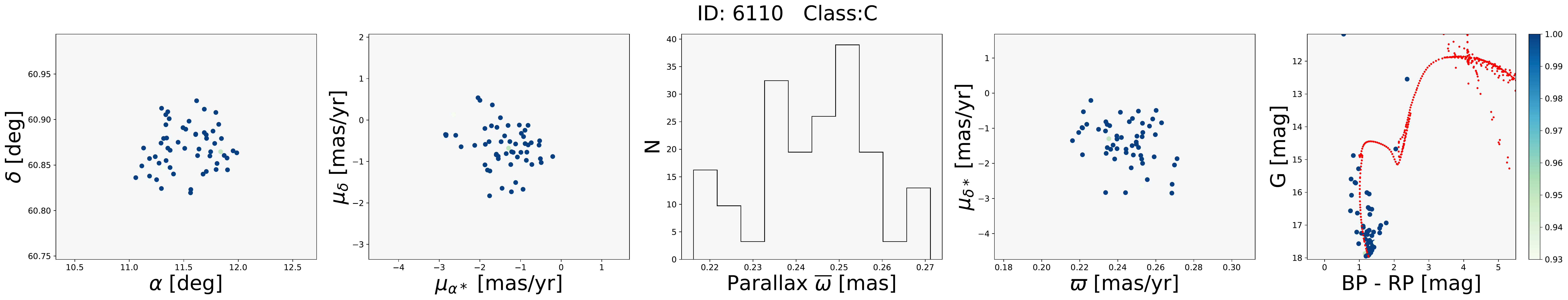}
    }
    \subfigure{
    \includegraphics[width=7in,height=1.5in]{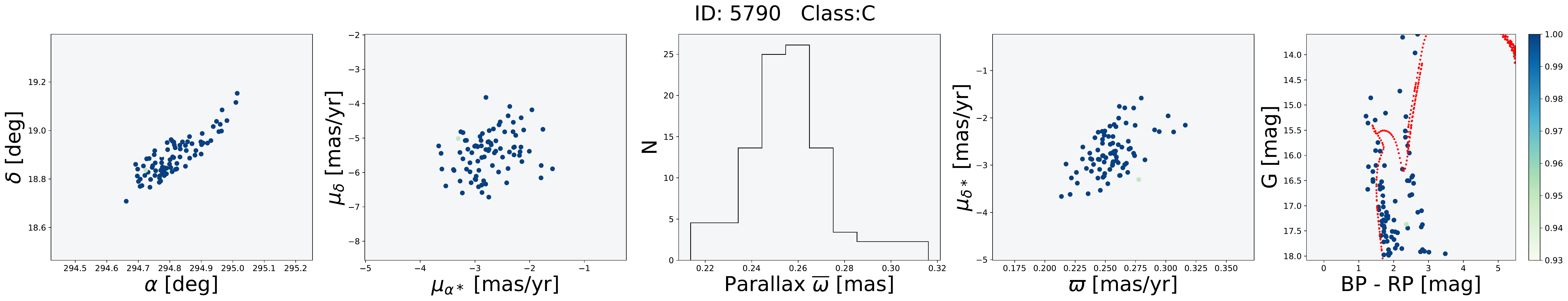}
    }
    \subfigure{
    \includegraphics[width=7in,height=1.5in]{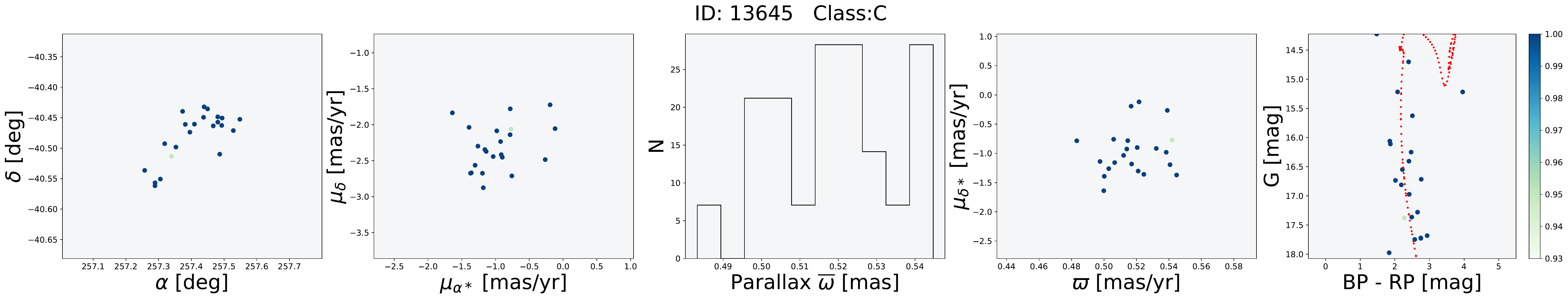}
    }
    \subfigure{
    \includegraphics[width=7in,height=1.5in]{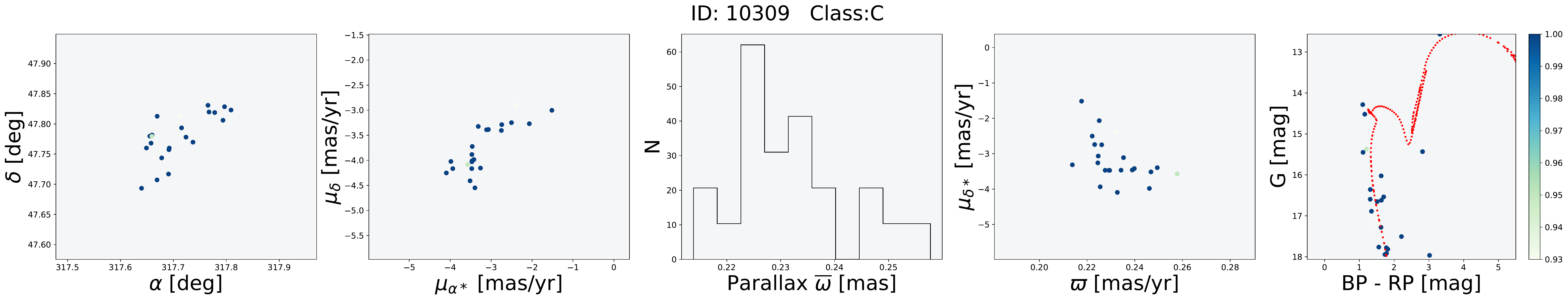}
    }
    \subfigure{
    \includegraphics[width=7in,height=1.5in]{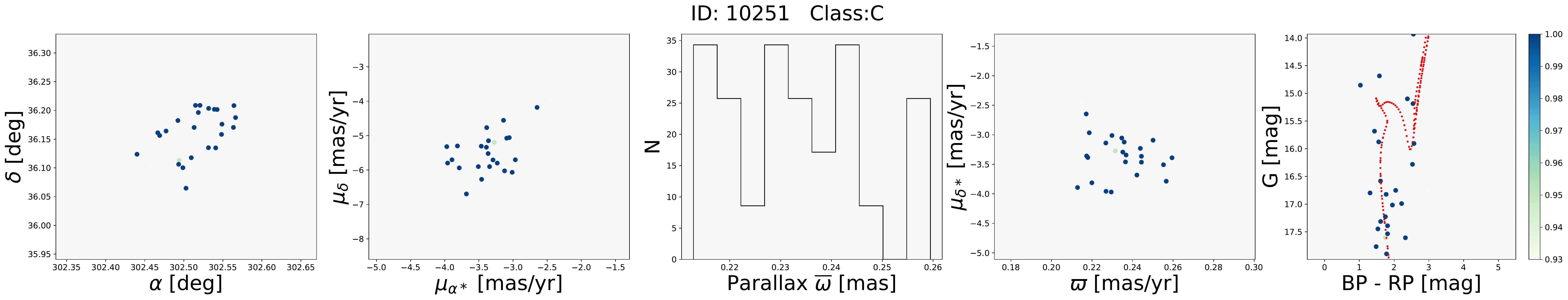}
    }
    \end{center}
    \caption{As same as Figure~\ref{fig:classA}, but for Examples of class C. }
    \label{fig:classC}
\end{figure*}

The OCs distribution of Class A in the celestial sphere is shown in Figure~\ref{fig:oc_distribution}.
Some examples of comparison of our results with previous work based on Gaia data are given in Figure~\ref{fig:hb}. Left to right: spatial distribution, parallax histogram, proper-motion distribution, color-magnitude diagrams (CMDs) with the best-fitting isochrone line.
The cross-matched candidates at High Galactic Latitude are listed in Figure~\ref{fig:hb}.
The candidates are divided into class A (see Figure~\ref{fig:classA}), class B (see Figure~\ref{fig:classB}), and class C (see Figure~\ref{fig:classC}), respectively.
From those figures, some member stars deviate from the main sequence, possibly due to inhomogeneous heavy reddening and/or field star contamination. 

In future work, the stars are further kicked out using a rational algorithm, such as the Bayesian algorithm, to authenticate class B and class C.
To determine if the clusters are real or not, just two fundamental parameters, age, and metallicity are inferred using the isochrone fitting method. 
However, more information about those OCs is expected to be detected by subtler methods and models, like the advanced stellar population synthesis model (ASPS)~\citep{Li2016A,2017RAA}.

\begin{table}[htbp]

    \caption{Parameters of 1,179 OCs identified in this work.}
    \label{tab:tab1}
    \begin{tabular*}{\hsize}{@{}@{\extracolsep{\fill}}lcccccccccccr@{}}
    
    \toprule
        ID & $N_{mem}$ & ra & ra\_sigma & dec &  ...& $N_{17}$& Z & age & N\_rv & RV\_mean & RV\_std&RV\_mad\\
         & &[deg]&[deg]&[deg]&...& &[$log (Z/ Z\odot)$]& [log(age/yr)] & & [km s$^{-1}$]& [km s$^{-1}$]& [km s$^{-1}$]\\

        \midrule
         34 & 335 & 97.3986 & 0.0018 & -31.2856 & ... & 200 & 0.44 & 8.630 & 16 & 60.096 & 2.422 & 1.679 \\ 
        36 & 46 & 262.9939 & 0.0061 & -67.0538 & ... & 16 & 0.44 & 8.720 & 1 & 10.215 & 0 &  \\ 
        41 & 60 & 76.904 & 0.0021 & 17.5737 & ... & 53 & 0.23 & 8.180 & 9 & 1.927 & 31.938 & 17.037 \\ 
        44 & 145 & 156.3305 & 0.2932 & -72.5262 & ... & 128 & -0.58 & 8.420 & 48 & -22.24 & 12.49 & 6.622 \\ 
        46 & 54 & 283.7608 & 0.0002 & -30.4795 & ... & 22 & -0.18 & 5.301 & 0 &  &  &  \\ 
        48 & 73 & 96.6488 & 0.001 & -9.6403 & ... & 58 & 0.2 & 8.180 & 12 & 38.94 & 3.455 & 2.819 \\ 
        53 & 45 & 277.8509 & 0.0002 & -32.3461 & ... & 32 & -0.58 & 6.146 & 0 &  &  &  \\ 
        57 & 38 & 147.2769 & 0.0165 & -65.2658 & ... & 32 & 0.16 & 7.730 & 6 & 5.265 & 13.063 & 10.939 \\ 
        58 & 31 & 332.9423 & 0.0049 & 45.1793 & ... & 27 & 0.44 & 8.630 & 4 & -50.106 & 4.352 & 3.254 \\ 
        59 & 631 & 48.6797 & 0.0156 & 47.2357 & ... & 476 & 0.27 & 8.090 & 68 & -29.14 & 22.522 & 8.719 \\ 
        60 & 347 & 105.127 & 0.0048 & -20.5784 & ... & 260 & -0.28 & 8.270 & 32 & 79.971 & 32.753 & 15.178 \\ 
        62 & 45 & 275.9221 & 0.0001 & -30.36 & ... & 27 & 0.16 & 8.240 & 0 &  &  &  \\ 
        64 & 105 & 93.0299 & 0.001 & 5.4479 & ... & 78 & -0.28 & 7.310 & 1 & 21.689 & 0 &  \\ 
        68 & 39 & 107.9666 & 0.0052 & -23.5535 & ... & 21 & 0.44 & 8.420 & 3 & 83.351 & 42.514 & 32.558 \\  
        ...&...&...&...&...&...&...&...&...&...&...&...&...\\
        \bottomrule
    \end{tabular*}
	\footnotesize
N$_{mem}$ is the number of cluster members, N$_{17}$ is the numbers of  magnitude less than 17 and N$_{rv}$ is the number of members that have radial velocity.
RV$_{mean}$ , RV$_{std}$ and RV$_{mad}$are the mean value, standard deviation and median value of RV dispersion, respectively.
The fields in the table are described in Table~\ref{tab:description_OC}.
This table and its description are available in their entirety in machine-readable form.
	
\end{table}

\begin{table*}[!ht]
    
    \centering
    \caption{Description of the catalog of star cluster properties (Table~\ref{tab:tab1}).}
    \label{tab:description_OC}
    \begin{tabular*}{\hsize}{@{}@{\extracolsep{\fill}}lcclccr@{}}
       \toprule 
       Column& Format& Unit& Description
 \\

\midrule 
ID&string&-&Cluster id in this work\\
N$_{mem}$&int&-&Number of star member\\
ra& float& deg& Mean right ascension of members\\
ra\_sigma& float& deg&Standard deviation of right ascension\\
dec& float& deg& Mean  declination of members of members\\
dec\_sigma& float& deg&Standard deviation of declination of members\\
l& float& deg& Mean  Galactic longitude of members\\
l\_sigma& float& deg&Standard deviation of  Galactic longitude\\
b& float& deg& Mean   Galactic latitude of members of members\\
b\_sigma& float& deg&Standard deviation of  Galactic latitude of members\\
plx& float& mag& Mean parallax of members\\
plx\_sigma& float& mag&Standard deviation of parallax\\
pmra& float& $mas ~yr^{-1}$& Mean proper motion in right ascension of members\\
pmra\_sigma& float&$ mas~ yr^{-1}$& Standard deviation of proper motion in right ascension\\
pmdec &float & $mas~ yr^{-1}$& Mean proper motion in declination of members\\
pmdec\_sigma& float& $mas~ yr^{-1}$& Standard deviation of proper motion in declination\\
n17&int&-& Numbers of magnitude less than 17 magnitude\\
Z&float& $\log (Z / Z \odot)$& Cluster metallicity determined by the isochrone fit\\
age&float& $\log (\text { age/yr })$& Cluster age determined by the isochrone fit\\
N\_rv&int& - &Number of radial velocity members\\
RV\_mean &float& $km~ s^{-1}$ &Mean radial velocity of members\\
RV\_std& float& $km ~s^{-1}$& Standard deviation of radial velocity\\
RV\_mad&float& $km ~s^{-1}$ &Median value of radial velocity dispersion of members\\
class&int&-& Class of OC in this work\\
\bottomrule
    \end{tabular*}

\end{table*}

\begin{table*}[!ht]
    
    \centering
    \caption{The Catalog of Cluster Members}
    \label{tab:sources}
    \begin{tabular*}{\hsize}{@{}@{\extracolsep{\fill}}lcccccr@{}}
       \toprule Gaia DR3 ID & $\alpha$ $(\mathrm{deg})$ & $\delta$ $(\mathrm{deg})$ &...& rv\_err&probs\_final & Cluster ID \\

\midrule
3430920657456186496&87.6612&26.9089&...&8.044&1.0&14154\\
 3430921653888613760&87.5773&26.923&...&7.76&1.0&14154\\
3430921787029123200&87.5869&26.9451&...& 12.972&1.0&14154\\
\bottomrule
    \end{tabular*}

    \begin{tablenotes}
    \item{
    This table and its description are available in their entirety in machine-readable form.}
    \end{tablenotes}
\end{table*}




\subsection{Linking length factor $b_{FoF}$}
In the study, we set the value of the linking length factor ($b_{FoF}$) to 0.2, in accordance with~\cite{Liu&Pang2019}. \cite{Liu&Pang2019} pointed out that they chose 0.2 because the value is commonly used in the dark matter halo identification of cosmological simulations~\citep{Springel2001}.

We are very concerned about whether different values of $b_{FoF}$ will affect OC identifications because we noticed that $b_{FoF}$ would significantly impact the clustering results.
To verify the reasonableness of taking 0.2 for $b_{FoF}$, we first selected the real star data of 5 regions from Gaia DR3, i.e., ID 1325 (106905 stars), ID 76 (107417 stars), ID 14 (64011 stars), ID 67 (82124 stars), and ID 18 (58714 stars). We then identified OC using different $b_{FoF}$ values.

\begin{figure}[htbp]
    \centering
    \includegraphics[width=4in,height=2.5in]{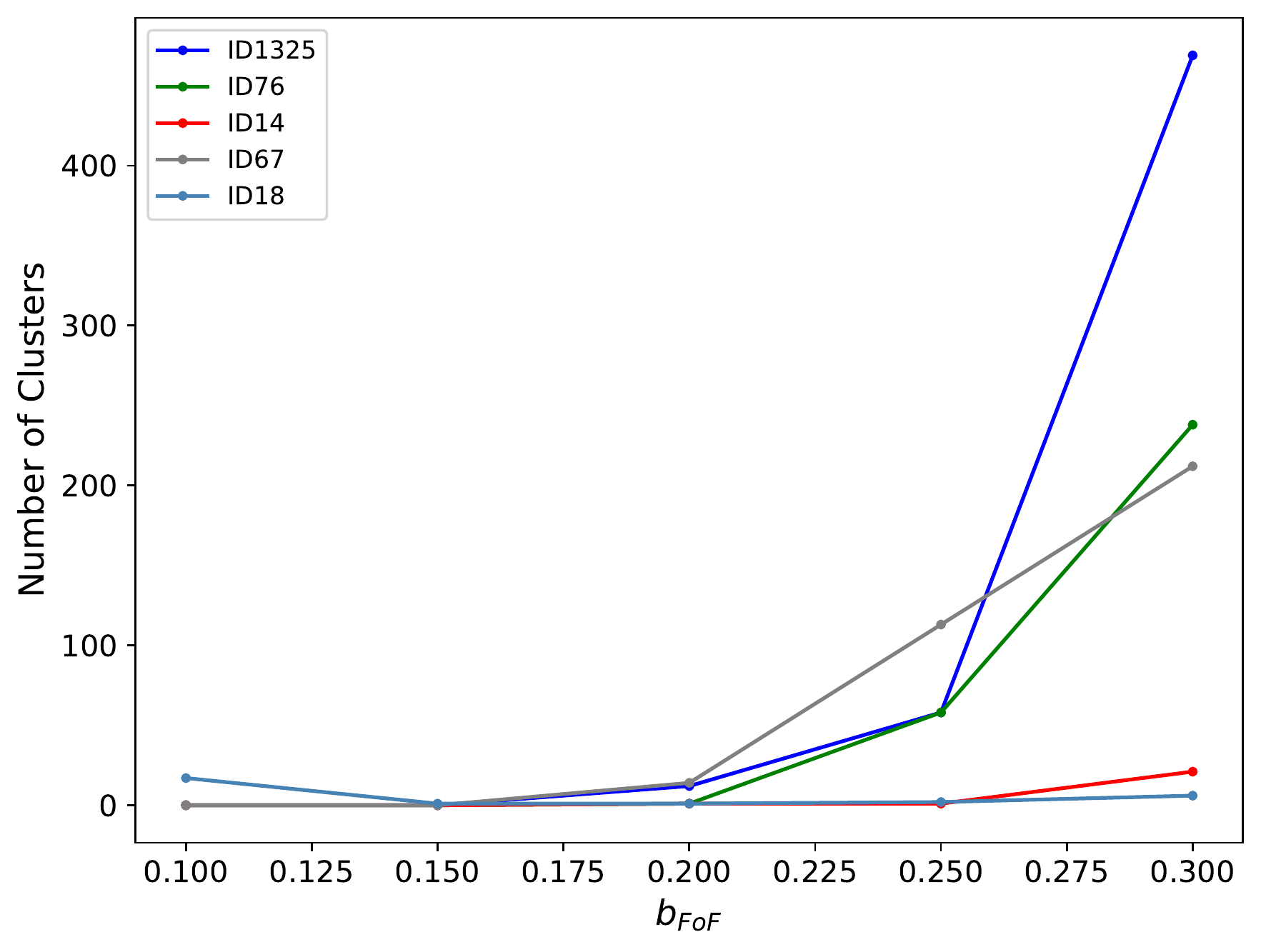}
    \caption{The comparison of different $b_{FoF}$ for 5 test datasets.}
    \label{fig:b_link}
\end{figure}

The experimental results (see Figure~\ref{fig:b_link}) show that the larger the $b_{FoF}$, the larger the number of detected clusters and the smaller the average cluster size. When $b_{FoF}$ is less than 0.2, the identified groups are basically the same. However, the identified groups increase significantly when $b_{FoF}$ is greater than 0.2.

Then, we selected the well-studied M67 cluster for testing. After a square query with a side of 5.5 degrees around an RA/Dec coordinate (132.85, 11.83) in Gaia DR3, we created a test dataset including 172557 stars of M67 (NGC2682). After the same data preprocessing, we obtained 42568 stars. We identified OCs using $b_{FoF}$ with 0.1, 0.2, and 0.3 and obtained 2, 10, and 60 groups, respectively. All different $b_{FoF}$ can identify M67 correctly.

According to the results of the above two experiments, the value of $b_{FoF}$ greatly influences the identification of OCs. However, considering that each candidate needs to be verified manually at a later stage, taking the value of $b_{FoF}$ as 0.2 may be a relatively reasonable compromise to balance the workload of manual verification and the correct rate of the identification model.


\section{Conclusions}
\label{summary}
To our latest knowledge, over 7,000 OC candidates have been found in our galaxy using different methods and algorithms. Identifying and confirming whether newly documented clusters in different published catalogs are genuine SCs requires a census of homogeneous member stars, which will be a challenging but necessary effort. 

We carried out a broad blind search for galactic star clusters. According to the probabilities, all candidates were divided into three classes, i.e., 1,194 (class A), 5,252 (class B), and 5,925 (Class C). 
After a series of stringent examinations, 1,179 true likely OCs in class A are present in this study. 

To sum up, this work enriches the OC sample of tracer galaxies within 5kpc nearby the Solar System, especially for the study of the local arm. This catalog will serve the community as a useful resource for tracing the chemical and dynamic evolution of the MW.

To determine if the clusters are real or not, we currently use the isochrone fitting method with two fundamental parameters, i.e., age and metallicity. 
Obviously, we need to develop new algorithms to discover more OCs accurately in the future.
One possible approach is to combine spectroscopic data from member stars to estimate more information about cluster parameters.
In addition, more than 10,000 objects in classes B and C still need to be identified using more advanced models. In addition, many candidate open clusters with complex main sequences require more advanced models for fitting and identification.
It is worth mentioning that binary open clusters are likely to exist in the true OCs we reported this time, which is worthy of further study.
There are some clusters that have tidal tails, such as ID00236, which could be disintegrating OCs.
Some are likely binary OC candidates that need to be identified further.
This is worthy of further investigation in the future.

\begin{acknowledgments}
This work is supported by the National SKA Program of China No 2020SKA0110300,  Joint Research Fund in Astronomy (U1831204) under cooperative agreement between the National Natural Science Foundation of China (NSFC) and the Chinese Academy of Sciences (CAS). Funds for International Cooperation and Exchange of the National Natural Science Foundation of China (11961141001). National Natural Science Foundation of China (No. 11863002), Yunnan Academician Workstation of Wang Jingxiu (202005AF150025),China Manned Space Project with NO.CMS-CSST-2021-A08 and Sino-German Cooperation Project (No. GZ 1284). Science and Technology Program of Guangzhou, China (2023A03J0016).
\end{acknowledgments}



\bibliography{references}{}
\bibliographystyle{aasjournal}

\end{sloppypar}
\end{CJK*}
\end{document}